\documentclass[onecolumn,showpacs,amsmath,amssymb,noshowpacs]
{revtex4-2}
\usepackage{amsfonts}
\usepackage{amsmath}

\usepackage{array}
\usepackage{braket}
\usepackage{enumerate}
\usepackage{titlesec, blindtext, color}
\usepackage{hyperref}
\usepackage{graphicx}

\usepackage[compat=1.1.0]{tikz-feynman}

\newcommand{\mz}[1]{{#1}}

\newcommand{\ms}[1]{{#1}}

\newcommand{\mzc}[1]{{#1}}

\begin{document}

\title{Non - renormalization of the fractional quantum Hall conductivity by interactions%
}
\author{M. Selch}
\email{maik.selch@t-online.de}
\author{M. A. Zubkov}
\email{mikhailzu@ariel.ac.il}

\author{Souvik Pramanik }
\email{souvick.in@gmail.com}

\author{M. Lewkowicz}
\email{lewkow@ariel.ac.il}

\date{\today }

\begin{abstract}
We investigate the theory of the fractional quantum Hall effect (QHE) proposed a long time ago by Lopez and 
Fradkin \cite{Fradkin1991chern} to describe the principal Jain series. The magnetic fluxes of the statistical gauge field attached to electrons remain at rest in the reference frame moving together with the electron liquid. In the laboratory reference frame the electric field of the statistical gauge field forms and screens the external electric field. The fractional QHE conductivity appears as a consequence of this screening already on the mean field theory level. We consider a relativistic extension of the model, and propose an alternative description of the fractional QHE based on macroscopic motion of the electron liquid within the Zubarev statistical operator approach. It is this macroscopic motion of electrons which in this pattern gives rise to the fractional QHE. Within this approach we propose the proof to all orders of perturbation theory that the interaction corrections can not change the above mentioned mean field theory result for the QHE conductivity.  
\end{abstract}

\maketitle
\tableofcontents

\affiliation{Ariel University, Ariel 40700, Israel}

\affiliation{Ariel University, Ariel 40700, Israel}

\affiliation{Physics Department, Ariel University, Ariel 40700, Israel}

%\author{Ruslan A.~Abramchuk}
%\email{abramchuk@phystech.edu}
%\affiliation{Physics Department, Ariel University, Ariel 40700, Israel}
%\affiliation{On leave of absence from Kurchatov Complex for Theoretical and Experimental Physics, B. Cheremushkinskaya 25, Moscow, 117259, Russia}

\affiliation{Physics Department, Ariel University, Ariel 40700, Israel}

\section{Introduction}

The quantum Hall effect (QHE) is one of the most intriguing and subtle phenomena in solid state physics.  It impacts both the theoretical understanding of the nature of electron liquids and applications to electronics. The QHE implies a quantization of the Hall resistance and occurs in two-dimensional solid state systems at low temperatures, in the presence of magnetic fields or nontrivial band topology. The integer and fractional QHE are distinguished. The former, first observed in \cite{klitzing1980quantized}, implies quantization of the Hall resistance in integer multiples of the von Klitzing constant $h/e^2$, where $h$ is
Planck's constant and $e$ is the elementary charge, while the quantization in terms of a rational number times the same constant is implied by the latter, as first observed in \cite{tsui1982two}. The fractional QHE arises in the case of fractional filling of Landau levels. It is a non - perturbative phenomenon associated with the appearance of emergent quasi-particles with fractional charges. \par

Building up a theory of the QHE began from the concept of Landau levels in the bulk and passed to the description of
edge states. 
The first explanation of the fractional QHE (FQHE) has been given utilizing the
concept of many - electron wavefunctions to describe correlated electron states leading to quasi-particles with fractional charges \cite{laughlin1983anomalous}. A further development of the theory led to the composite
fermion theory, which is able to explain the broad range
of filling fractions observed experimentally \cite{jain1989composite}. A field theory description in terms of Chern - Simons theory has been given in \cite
{Zhang1989effective} and \cite{Fradkin1991chern}. The analysis of disorder effects on the QHE brought about a twist in the understanding of the QHE as a purely  bulk phenomenon. The localization of electrons in the bulk due to disorder pushes the quantum Hall current towards the boundaries of the samples, and results in the importance of the role of edge currents. The edge theory of the FQHE has been developed in \cite{halperin1982quantized}, \cite{Wen1990chiral} as well as \cite{Kane1995impurity}. \par

Recently the development of technology enhanced the study of the FQHE within
various materials. Samples of novel materials were grown, where complicated fractional states are realized \cite{shayegan1998quantum}. The appearance of new filling fractions at higher Landau levels was proposed for ultraclean samples of graphene \cite{dean2011multicomponent}. In addition to the conventional two-dimensional solid state systems, other systems also host the QHE such as topological insulators \cite{hsieh2009observation} and photonic
systems \cite{lu2014topological}. Materials that possess the QHE appear in ultra-low-power transistors \cite{sarama1997perspectives}. The fractional QHE has been discussed for bilayer systems, in which the interlayer interaction results in the appearance of novel quantum
states \cite{eisenstein2004bose}. Unique properties of graphene admit the observation of the QHE at room temperature, which may potentially lead to practical
applications in electronics \cite{novoselov2005two}.\par

The Wigner - Weyl formalism represents a way to conveniently describe the QHE theoretically. It can be applied to both lattice and continuum models in order to express the quantum Hall current through topological invariants in phase space. These invariants incorporate Wigner - transformed Green functions \cite{Zubkov2020}. Within the Wigner - Weyl formalism an expression for the Weyl symbol of the Dirac operator is derived and the Groenewold equation solved in order to determine the Wigner - transformed Green function \cite{Suleymanov2021}. The Hall conductivity in interacting 2 + 1 D topological insulators and 3 + 1 D Weyl semimetals can be expressed using the complete
two - point Green function, similar to the non - interacting case \cite{Zhang2021}. Namely, the quantum Hall conductivity in the presence of interactions is shown to be given by the same topological expression as for the non - interacting system, but with the non-interacting
Green function replaced by the complete two - point Green function \cite{Zhang2019}. Later on, a generalization of the QHE to the case of non - uniform external 
fields was proposed. In such systems the topological invariant in phase space associated with the quantum Hall conductivity is expressed through the Weyl symbols of the two - point Green function \cite{Fialkovsky2020} and Moyal products. As expected, the QHE in the presence of arbitrary electron
interactions was shown to be determined completely by the filling fraction of Landau levels averaged over the ground state, valid for both integer and fractional effects \cite{Wu2020}. The Hall conductivity for the integer QHE (IQHE) in homogeneous systems in the presence of interactions in Harper representation is given as well by a topological invariant based on the Green function \cite{Selch2023}. The QHE in artificial systems may be described using a specific version of Wigner - Weyl calculus designed for lattice models with large values of magnetic flux through the lattice cells \cite{Zubkov2023,Chobanyan2024}. The anomalous QHE (AQHE), whose occurrence does not necessitate an external magnetic field, has been described to show that the conductivity takes fractional values times the von Klitzing constant \cite{Wu2021}. Non - equilibrium scenarios can be described using the Keldysh technique,
which again implies that the Hall conductivity may be expressed through Wigner - transformed Green functions \cite{Banerjee2021}. To describe the effect of interactions on the Hall conductivity, the non - interacting two - point Green function is replaced by the complete interacting Green function \cite{Zhang2022}. 
 
Notice that a similar approach based on the Wigner - Weyl calculus 
can be followed to study other non - dissipative transport phenomena, including the
Chiral Magnetic Effect and the Chiral Separation Effect \cite{Zubkov2023b}.
Certain extensions of this approach that utilize multi - particle Green functions have been constructed as well to obtain a topological
expression for the Hall conductivity (valid for both ordinary QHE and AQHE) in systems with interacting electrons, external magnetic fields and non - trivial
band topology \cite{Miller2022b}. 

Both IQHE and FQHE are important for a purely theoretical understanding of topological phases of matter. The topological phase is a novel kind of quantum
state that cannot be described solely via a study of symmetry breaking \cite{wen1990topological}. Topological invariants, such as the Chern number, are used to 
classify such states inside topological insulators \cite{hasan2010topological}.

The theory of the FQHE proposed by Fradkin and collaborators supplements the non - relativistic action of electrons in an external magnetic field by a Chern - Simons term. The fermionic theory of the fractional QHE by Lopez and Fradkin \cite{Fradkin1991chern} is of particular interest for the present paper. In this theory the Chern - Simons term is introduced to describe an emergent statistical gauge field. It describes the general idea of composite fermions \cite{jain1989composite} within the particluar model, which are formed by attachment of magnetic fluxes to the electronic quasiparticles. An effective magnetic field appears  $B_{eff} = B-\rho/\theta$ as a result of screening of the external magnetic field $B$ by the statistical gauge field. Here $\theta$ is the Chern - Simons coupling, while $\rho$ is the electron density. A similar screening occurs also for the electric field \cite{simon1998chern}. The original electron and composite fermion descriptions are claimed to be equivalent for the values $\theta = \frac{1}{2\pi 2s}$ and any positive integer $s$. A projection of the original theory of electrons in a magnetic field onto the lowest Landau level, which we consider in this work, implies an approximate particle hole symmetry which may be implemented naturally in a relativistic framework of Dirac composite fermions as first elaborated in \cite{son2015composite} and reviewed in \cite{son2018dirac}. The relativistic extension has been considered as a useful tool for the further development of the bulk theory. The corresponding Dirac composite fermion theory is detailed in \cite{nguyen2018particle,Nguyen:2021vzp}. In the present paper we rely on a different relativistic extension of the model proposed by Lopez and Fradkin. Our purpose is to give the proof of the non - renormalization of the FQHE by interactions in a way similar to what was done earlier for the IQHE in \cite{Zhang2022}. Besides, this relativistic extension will allow us to use the formalism of the Zubarev statistical operator developed earlier for the truely relativistic systems. In \cite{selch2024effective} the Zubarev statistical operator has been represented in the form of a functional integral over the quantum fields. 
%We follow this representation, but introduce to it an extra ingredient - a velocity $u$ of macroscopic motion appears as a dynamical variable. 
It appears that there exists an exceptional case when the macroscopic velocity is orthogonal to external electric field with its absolute value given by the ratio $E/B$. In this way we arrive at the description of the electron liquid moving with this velocity and the resulting electric current is that of the fractional quantum Hall effect.\par

The paper is organized as follows. In Section \ref{SectPrel} we introduce notations and describe the mean field approximation in the model by Lopez and Fradkin of \cite{Fradkin1991chern}. In Sect. \ref{SectRel} we propose the relativistic extension of the model. Notice, that this extension differs conceptually from that of \cite{nguyen2018particle}. Besides, we  propose a way to look at the problem based on the Zubarev statistical operator ans its representation through the functional integral. In Sect. \ref{SectInt} we give the proof that the QHE conductivity in the relativistic model under consideration remains unaffected by interactions. In Sect. \ref{SectConcl} we end with conclusions. 

\section{ Preliminaries}
\label{SectPrel}
\subsection{Electric current for the electrons with Coulomb interactions}

\label{Sect5}

We start from the expression for the partition function of non - relativistic electrons with mass $M$ interacting by Coulomb forces
\begin{equation}
	Z = \int D\psi D\bar{\psi} D \lambda e^{
		iS[\lambda] + i\int d^3 x \bar{\psi} (\hat{Q}[\lambda]) \psi}
	\label{Z0}
\end{equation}
with  
\begin{eqnarray}
	\hat{Q}[\lambda] &=&  i\partial_0 - A_0(z)    +
	\mu - \lambda(z)+ \frac{1}{2M} \vec{D}^2  \nonumber \\
	S & = & \mz{\frac{1}{2} \int d^3 z d^3 \lambda(z) [V^{-1}](z,z^{\prime })
		\lambda(z^\prime)}  \nonumber \\
	\vec{D} & = & \vec{\partial} - i \vec{A} \label{ZLF0}
\end{eqnarray}
Here $\lambda$ is the dynamical field of electric potential, 
$V$ is the Coulomb potential, \mz{$V^{-1} = \frac{1}{e^2}\Delta^{(3)}$  is operator inverse to $V$, while $[V^{-1}](z,z') = \bra{z}V^{-1}\ket{z'}$ is its matrix elements. Here $\Delta^{(3)}$ is the three - dimensional Laplacian. Throughout the paper we use the relativistic units with $\hbar = c = 1$. Moreover, we include $|e|$ in the definition of Electric field and in the definition of electric current. As a result, for example, the Hall conductivity calculated below in the conventional units should be multiplied by $e^2/\hbar$. }

Electric current is
then given by the variation of $Z$ with respect to the external gauge field: 
{\begin{equation}
	j^k(x) = i \frac{\delta Z}{Z \delta A_k} = -\frac{1}{Z}\int D\psi D\bar{\psi} D\lambda e^{iS[\lambda] + i\int d^3 x \bar{\psi} \hat{Q}[\lambda] \psi} \bar{\psi}(x) \frac{\partial}{\partial A_k}\hat{Q}[\lambda] \psi(x) \label{j00}
\end{equation}
In the zeroth order of perturbation theory, when $\lambda$ is set to zero (see \cite{Zubkov2020})  \begin{equation}
	j^k = \mzc{i} \mathrm{Tr}\, \int \frac{d^3 p}{(2\pi)^3} {G}^{(0)}_W(x,p)
	\partial_{p^k} Q^{(0)}_W(x,p),
	\label{electriccurrent0}
\end{equation}
where $\hat{Q}^{(0)} \equiv \hat{Q}[0]$ is the bare Dirac operator with interactions disregarded, $\hat{G}^{(0)} = [\hat{Q}^{(0)}]^{-1}$ is the non - interacting Green function \mz{(notice that we do not include $i$ in the definition of the Green function in Minkowski space - time)}. Eq. (\ref{electriccurrent0}) contains the Weyl symbols of these operators denoted by subscript $W$. Namely, let $\hat{O}$ be an operator with
	configuration space matrix element $O(x,y)=\langle x|\hat{O}|y\rangle $. The
	Weyl symbol of $\hat{O}$ is then defined by 
	\begin{align}
		O_W(x,p)=\int d^3rO(x+\frac{r}{2},x-\frac{r}{2})e^{-ipr}.\label{Weyl}
\end{align}
Derivation of Eq. (\ref{electriccurrent0}) for electric current in terms of the Wigner transformed quantities may be found, for example, in \cite{Zhang2022} and references therein.

In the following by bold letter $\hat{\mathbf{G}}$ we denote the interacting Feynman Green function, while by ordinary letter $\hat{Q}$ we denote the non -
interacting Dirac operator. Correspondingly, $\hat{\mathbf{Q}}$ is inverse to $\hat{\mathbf{G}}$ while  $\hat{G}$ is inverse to  $\hat{Q}$. 
We will throughout this work employ the subscript $W$ to denote the Weyl symbols of the corresponding operators. \mz{The definition of the Weyl symbol is given above in Eq. (\ref{Weyl}).} 
\textit{One can see that the electric current is to be calculated as the non
- renormalized velocity operator $\partial_{p^k} Q_W(x,p)$ averaged with the
aid of the interacting Green function $\mathbf{G}_W(x,p)$ (see, for example, \cite{Zhang2019,Zhang2022})  :}
 \begin{equation}
	j^k = \mzc{i} \mathrm{Tr}\, \int \frac{d^3 p}{(2\pi)^3} \mathbf{G}_W(x,p)
	\partial_{p^k} Q_W(x,p)
	\label{electriccurrent}
\end{equation}
 It has been
proven in \cite{Zhang2022} that when the interactions are taken into account
perturbatively, the non - renormalized velocity operator may be replaced by
the renormalized one $\partial_{p^k} \mathbf{Q}_W(x,p)$, where $\mathbf{Q}%
_W(x,p)$ obeys the Groenewold equation  
\begin{equation}
\mathbf{Q}_W(x,p) \star \mathbf{G}_W(x,p) = 1
\end{equation}
and $\star$ denotes the Moyal star product. \mz{It has been shown in \cite{Zhang2022} that the value of the QHE
conductivity remains the same as without
interactions if we take into account the latter perturbatively. This
observation brings us to the conclusion that the fractional quantum Hall effect
is a non - perturbative phenomenon. Below we will see that introduction of a statistical gauge field proposed by Lopez and Fradkin allows us to account for this non - perturbative phenomenon already on the level of  mean field theory. We will further see using the relativistic extension of the model that the interaction corrections to this mean field result vanish perturbatively. }  

\subsection{The model by Lopez and Fradkin and the mean field approximation}

\mz{Lopez and Fradkin \cite{Fradkin1991chern} argued that the standard model of electrons interacting by Coulomb forces is equivalent to their model defined by the partition function }
\begin{equation}
Z = \int D\psi D\bar{\psi} D \lambda D\mathcal{A}  e^{iS_{CS}[\mathcal{A}] +
iS[\lambda] + i\int d^3 x \bar{\psi} (\hat{Q}[\mathcal{A},\lambda]) \psi}
\label{Z1}
\end{equation}
with  
\begin{eqnarray}
\hat{Q}[\mathcal{A},\lambda] &=&  i\partial_0 - A_0(z) -  \mathcal{A}_0 +
\mu - \lambda(z)+ \frac{1}{2M} \vec{D}^2  \nonumber \\
S & = & \mz{\frac{1}{2} \int d^3 z d^3 \lambda(z) [V^{-1}](z,z^{\prime })
\lambda(z^\prime)}  \nonumber \\
S_{CS} & = & \frac{ \theta}{4} \int d^3 z \epsilon^{ijk} \mathcal{A}_i 
\mathcal{F}_{jk}  \nonumber \\
\vec{D} & = & \vec{\partial} - i \vec{A} - i \vec{\mathcal{A}}\label{ZLF}
\end{eqnarray}
$V$ is the Coulomb potential, $V^{-1}$ is the operator inverse to $V$, while $[V^{-1}](z,z') = \bra{z}V^{-1}\ket{z'}$ is its matrix element. \ms{$S_{CS}$ is the Chern-Simons action comprising a so-called statistical gauge field which moreover appears in the covariant derivative of the electrons. The Chern-Simons term is topological. It does contribute neither to the energy nor the momentum carried by the fields. Lopez and Fradkin integrated out the fermions and the statistical gauge field in their model in an RPA approximation. The residual effective action contains a Chern-Simons term comprising the external electromagnetic gauge field sourced exactly by $S_{CS}$ as given above in terms of the statistical gauge field. The Chern-Simons effective action is the general effective field theory manifestation of the QHE. The strength of the topological term is determined by $\theta = 1/(2\pi \,2s)$, and $s$ is an integer. The Chern-Simons term attaches $2s$ quanta of magnetic flux (with magnetic flux quantum $\Phi_0=2\pi$ in our units) to each electron. This gives rise to a new effective quasiparticle, the so-called composite fermion. It is subjected to an effective electromagnetic field, the sum of the external field plus the screening statistical field. The former is not confined to the sample, while the latter is. The advantage of the composite fermion formulation is the existence of a mean field theory around which we may expand meaningfully. The flux attachment procedure which maps a FQHE for electrons to an IQHE for composite fermions. The requirement for $s$ to be an integer will be explained in detail in Appendix \ref{AppDual}. Most importantly its value fixes the statistics of the quasiparticles. We intend to continue working with fermions. Other choices of $s$ would lead to Abelian anyons as quasiparticles.}\par
The arguments presented by Lopez and Fradkin in favor of the mentioned equivalence are not based on the conventional
perturbation expansion. Instead, the representation of the theory through the sum over the worldlines of particles is used. \mz{It appears that the partition function of Eq. (\ref{ZLF}) equals the partition function given by the same expression, from which the statistical gauge field is removed completely. This has been proven also in Appendix \ref{AppDual} using duality transformation. }

{The mean field theory results in equations (we denote by $\bar{\rho}$ the average particle density)
	\begin{eqnarray}
		&& \frac{\theta}{2}  \epsilon^{ijk}  {\cal F}_{jk} = J^i \nonumber\\ &&
		J^0(z) = \int d^3 z'  [V^{-1}(z,z')] \lambda(z') +  \bar{\rho}(z)\nonumber\\&& \vec{J} = \langle(\bar{\psi} \psi, \bar{\psi} \frac{-i \vec{D}}{M} \psi)\rangle\label{meanfield0}
	\end{eqnarray}
	An obvious solution of these equations is
	\begin{eqnarray}
		&& {\theta} {\cal B} = -\bar{\rho} \nonumber\\&& {\theta} {\cal E}^i =  \epsilon^{ij}\bar{j}^j \nonumber\\ &&
		\lambda(z) = 0 \nonumber\\&& \bar{\rho} = \langle\bar{\psi} \psi \rangle
		\nonumber\\&& \bar{j}^k = \langle\bar{\psi}\frac{-i D^k}{M} \psi \rangle
		\end{eqnarray}}

The magnetic field $B$ is screened by the statistical gauge field giving rise to the effective magnetic field $B_{eff}$. If precisely $p$ Landau levels of $B_{eff}$ are occupied, then we have $\bar{\rho} = \frac{p}{2\pi}B_{eff}$. Therefore, we arrive at the equation for the determination of $B_{eff}$:
$$
B_{eff} = B - \frac{\bar{\rho}}{\theta} = B - 2sp B_{eff}
$$
and $B_{eff} = \frac{B}{1+2sp}$.
A similar situation takes place for the effective electric field $$
{E}_{eff}={E} - j_{QHE}/\theta = {E} - 2\pi \, 2s\, j_{QHE},
$$ 
where $j_{QHE}$ is the QHE current, which  according to Appendix \ref{app0} is given by 
\begin{equation}
 j_{QHE}  \approx  \frac{1}{2\pi}\mathcal{N} E_{eff} = \frac{1}{2\pi}\mathcal{N} ({E} - 2\pi \, 2s j_{QHE})\,.
\end{equation}
The average Hall conductivity at zero temperature is then given by $\sigma_{xy} = 1/(2 \pi)\,\times\, \frac{\cal N}{1+2s {\cal N}}$
 where $\mathcal{N}$ is an integer number. Using the machinery developed in \cite{Zubkov2020} we represent it as  the following 
expression (to be calculated after Wick rotation using Euclidean (Matsubara) Green functions) 
\begin{eqnarray}
\mathcal{N} & = & {\frac{T}{3! 4\pi^2 \mathcal{S}}} \, \int \,\mathrm{Tr}\,
d^3p \, d^3x\, \epsilon_{ijk}\, \Big[ (G_{eff})_W(p,x)\ast \frac{\partial
(Q_{eff})_W(p,x)}{\partial p_i}  \nonumber \\
&& \ast (G_{eff})_W(p,x)\ast \frac{\partial (Q_{eff})_W(p,x)}{\partial p_j}
\ast (G_{eff})_W(p,x)\ast \frac{\partial (Q_{eff})_W(p,x)}{\partial p_k} %
\Big]\,.  \label{calM4}
\end{eqnarray}
where $(Q_{eff})_W (p,x)$ is the Wigner transform of $Q_{eff} (p,x)$
respectively, with  
\[
Q_{eff}(p,x) = i\omega - H_{eff}(p_x,p_y - B_{eff} x).
\]
and $(G_{eff})_W$ is the Green's function calculated by inversion of $%
(Q_{eff})_W$. Here, in the mean field approximation the effect of
interactions is included through the effective magnetic field $B_{eff} = B -
\rho_0 / \theta$ \cite{Fradkin1991chern}. Direct calculation gives $\mathcal{N}
= p$ (for the details see Appendixes \ref{app1} and \ref{app2}) and we arrive at the fractional QHE
\begin{equation}
	\sigma_{xy} = 1/(2 \pi)\,\times\, \frac{p}{1+2sp}
\end{equation}
One can see that the screening of external electric field by the statistical electric field is crucial for the appearance of the fractional QHE. \mz{This reasoning can also be found in \cite{simon1998chern}, Sect. 3.1.} A qualitative explanation of this screening will become clear below, when we will discuss the transformations between different reference frames. Namely, the screening statistical magnetic field appears in the reference frame moving together with the electron liquid. In that reference frame it represents the magnetic fluxes attached to the electrons. Transformation to the laboratory reference frame (where the external electric field is nonzero, and the macroscopic velocity of the electron liquid is nonzero as well) results in the Lorentz transformation of the statistical gauge field fluxes. This is how the screening statistical electric field appears. \mz{In particular, the result of this screening is the appearance of an effective fractional value of electric charge for the electronic quasiparticles $e' = \frac{e}{1+2sp}$. }

\section{From non - relativistic to relativistic fermion action}

\label{SectRel}

\subsection{Reduction of relativistic model to the non - relativistic model of Fradkin and Lopez}

{Let us consider the original model by Lopez and Fradkin in Minkowski space - time. Its Lagrangian is
\begin{align}
	\mathcal{L}=i\psi^{\dagger}(\partial_t+iA_0+i\mathcal{A}_0-i\mu^{NR})\psi -\frac{1}{2m^{NR}}[(\partial_i+iA_i+i \mathcal{A}_i)\psi ]^\dagger [(\partial_i+iA_i+i \mathcal{A}_i)\psi ] +\frac{\epsilon^{\mu\nu\lambda}}{\mz{8\pi s} }\mathcal{A}_{\mu}\partial_{\nu}\mathcal{A}_{\lambda}+\mathcal{L}_{int}^{NR}.
	\label{nonrellagrangian}
\end{align}
(Notice that spatial components of $A_\mu$ differ by sign from $\vec{A}$, and the same for $\cal A$.) Here $ \mathcal{L}_{int}^{NR}$ is the interaction term (it contains the interaction of $\psi$ with the field $\lambda$ and the action for $\lambda$ itself).
At {$\mu$} slightly above {$M$} the given \mz{theory} appears as the non - relativistic limit of the theory with the $2+1$ D Lagrangian of the electron gas 
\begin{align}
	\mathcal{L}=i\overline{\psi}\gamma^{\mu}(\partial_{\mu}+iA_{\mu}+i\mathcal{A}_\mu + \mz{i 
	\lambda_\mu})\psi -M\overline{\psi}\psi +\mu \overline{\psi}\gamma^0\psi +\frac{\epsilon^{\mu\nu\lambda}}{8\pi s}\mathcal{A}_{\mu}\partial_{\nu}\mathcal{A}_{\lambda}
	\label{nonsonlagrangian}
\end{align}
with $m^{NR}=|M|$ and $\mu =|M|+\mu^{NR}$. Within this relativistic Lagrangian $\psi$ is a $2+1$ D two - component Dirac spinor, while the $\gamma$ - matrices are actually proportonal to the Pauli matrices. Notice that this relativistic extension of the theory differs essentially from that of \cite{son2018dirac,nguyen2018particle}. Our extension is devoted to the consideration of the principal Jain sequence while, say, \cite{son2018dirac} was devoted specifically to the filling fraction $1/2$.  \mz{The lagrangian for the field $\lambda$ is now $3+1$ dimensional:
\begin{eqnarray}
	\mathcal{L}_{int}= - \frac{1}{4 e^2} (\partial_{[\mu} (A + \lambda)_{\nu]})^2 + L_{gauge\, fixing}[A+\lambda]
\end{eqnarray}
Here the fluctuations of the electromagnetic field around the external gauge field $A$ are denoted by $\lambda_\mu$ and contain $4$ components. Only three of these four components enter the interaction term with the two - dimensional electron gas ($\mu = 0,1,2$), but all four components enter the pure electromagnetic field action, which results in the conventional $2+1$ D propagator of $\lambda$.} In the non - relativistic limit only the zeroth component $\lambda_0$ is relevant, and it becomes the field $\lambda$ of the non - relativistic theory. The gauge fixing term $L_{gauge\, fixing}[A+\lambda]$ is to be added in order to make it possible to calculate propagators of the field $\lambda$ and perform perturbative calculations. We do not concretize this term here.   

The relativistic invariance is broken explicitly by the term containing the chemical potential.  Notice that in the "relativistic" model of Eq. (\ref{nonsonlagrangian}) we assume the regularization that subtracts the contribution of the Dirac sea from Hall current. As such a regularization we can choose the lattice regularization, in which the contributions (to the Hall conductivity) of some of the Landau levels situated far below zero are proportional to large negative values of Chern numbers, which cancel precisely the positive contributions of the remaining Landau Levels situated below zero. In Pauli - Villars regularization the same job is done by the Pauli - Villars regulator. {This becomes especially obvious if we neglect interactions. Then for the calculation of electric current we can choose one  Pauli - Villars regulator with large but finite mass. For positive values of  $\mu$ the contributions (to the Hall current) of the regulator's Landau levels eith negative energies are precisely equal to the similar contributions of the original electron field. This occurs because these contributions do not depend on mass. We subtract the contributions of the Pauli - Villars regulator from the contributions of the original electron field. As a result the cancellation occurs. At the same time there are no contributions to the electric current of the positive Landau levels of the regulator as long as the mass of the regulator is larger than $\mu$. That's why we are left with the contributions of positive Landau levels to the electric current due to the original electron field, while the Dirac sea should be disregarded. }

{Let us consider the situation with the external electric field $E$ directed along the $x$-axis, and a nonzero magnetic field. We choose the gauge setting $A_0 = -Ex,\, A_y = B x, \,A_x = 0$. The partition function of the new "relativistic" model in Minkowski space - time is given by 
	\begin{equation}
		Z = \int D\psi D\bar{\psi} D \lambda D{\cal A} e^{i S_{CS}[{\cal A}] + iS_g[\lambda]  + i\int d^3 x \bar{\psi} (\hat{Q}[{\cal A},A,\lambda]) \psi} 
	\end{equation}
	with 
	\begin{eqnarray}
		S_g & = &  - \frac{1}{4 e^2}\mz{\int d^4z} (\partial_{[\mu} (A + \lambda)_{\nu]})^2 +\mz{\int d^4 x} L_{gauge\, fixing}[A+\lambda]\nonumber\\ 
		S_{CS} & = &  \frac{\theta}{4} \int d^3 z \epsilon^{ijk} {\cal A}_i {\cal F}_{jk} \nonumber\\ 
		{D} & = & {\partial} + i {A} + i {\cal A} {+i\lambda}.  
	\end{eqnarray}
Let us denote by $\bar{J}^{\mu} = (\bar{\rho},\bar{j}^x,\bar{j}^{y})$  the average electric current of the quantum Hall system with $u = (1,0,0)$ in the {\it given} reference frame (with nonzero $E$). $\bar{J}^{\mu}$ is to be determined by equations of motion. }

{The Dirac operator is
\begin{eqnarray}
	\hat{Q}[{\cal A},\lambda] &=& i \gamma^0 \partial_t -\gamma^0 ({\cal A}_t - Ex) + \mu\gamma^{\nu} u_{\nu} - \gamma^{\mu}\lambda_\mu(z)+ \gamma^k  \Bigl( i {\partial}_k + Bx\hat{y}_k - {\cal A}_k\Bigr) - M  
\end{eqnarray}
The mean field theory results in equations
\begin{eqnarray}
	&& \frac{1}{2}  \epsilon^{ijk}  \theta{\cal F}_{jk} = J^i \nonumber\\ &&
	J^\mu(z) = \int d^3 z'  [D^{-1}(z,z')]^{\mu\nu} \lambda_\nu(z') + \overline{J}^\mu(z)\nonumber\\&& J^\mu = \langle \bar{\psi}\gamma^\mu \psi \rangle\label{meanfield1}
\end{eqnarray}
Here $D$ is the photon propagator in the chosen gauge. The particular case of Coulomb gauge leads to the second equation of Eq. (\ref{meanfield0})  in the non - relativisic limit.
An obvious solution of these equations in mean field approximation is
\begin{eqnarray}
	&& {\theta} {\cal B} = -\bar{\rho} \nonumber\\&& {\theta} {\cal E}^i =  \epsilon^{ij}\bar{j}^j \nonumber\\ &&
	\lambda(z) = 0 \nonumber\\&& \bar{\rho} = \langle\bar{\psi}\gamma^0 \psi \rangle
	\nonumber\\&& \bar{j}^k = \langle\bar{\psi}\gamma^k \psi \rangle
\end{eqnarray}}

{If we neglect  interactions (both with the fluctuations
of $\lambda$ and $\mathcal{A}$ over the mean field values) in the model of
Eq. (\ref{nonsonlagrangian}) the QHE conductivity may be calculated easily.
It is given (as well as in the non - relativistic model itself) by an
integer number ${\cal N} = p$ times the inverse von Klitzing constant divided by $1+2sp$. As it should, this is the original result obtained by Lopez and Fradkin for their non - relativistic model.    }

%%%%%%%%%%%%%%%%%%%%%%%%%%%%%%%%%%%%%%%%%%%%%%%%%%%%%%%%%%%%%%%%%%%%%%%%

%\section{Description of macroscopic motion of the electron liquid by the Zubarev statistical operator}

\subsection{Lorentz transformation}

{As above, we consider the situation with the electric field $E$ directed along the $x$-axis with a nonzero magnetic field. We choose the gauge $A_0 = -Ex, A_y = B x, A_x = 0$. The partition function in Minkowski space is given by 
	\begin{equation}
		Z = \int D\psi D\bar{\psi} D \lambda D{\cal A}  e^{i S_{CS}[{\cal A}] + iS_g[\lambda]  + i\int d^3 x \bar{\psi} (\hat{Q}[{\cal A},\lambda]) \psi} 
	\end{equation}
	with 
	\begin{eqnarray}
		S_g & = &  - \frac{1}{4 e^2}\mz{\int d^4 x} (\partial_{[\mu} (A + \lambda)_{\nu]})^2 +\mz{\int d^4 x} L_{gauge\, fixing}[A+\lambda]\nonumber\\ 
		S_{CS} & = &  \frac{\theta}{4} \int d^3 z \epsilon^{ijk} {\cal A}_i {\cal F}_{jk} \nonumber\\ 
		{D} & = & {\partial} + i {A} + i {\cal A}{+i\lambda}   
	\end{eqnarray}}

In the relativistic case, the Dirac operator is
	\begin{eqnarray}
		\hat{Q}[{\cal A},\lambda] &=& i \gamma^0 \partial_t -\gamma^0 ({\cal A}_t - Ex) + \mu\gamma^{\nu} u_{\nu} - \gamma^{\mu} \lambda_\mu(z)+ \gamma^k  \Bigl( i {\partial}_k + Bx\hat{y}_k - {\cal A}_k\Bigr) - M  
	\end{eqnarray}
where $u = (1,0,0)$ in the {\it given} reference frame {(the laboratory frame with nonzero $E$)}, $\hat{y}_k$ is the unit vector directed along the $k$ - th axis.
The mean field theory results in 
 ${\cal A}_t = -\frac{j_{QHE}}{\theta} x$ and $\vec{A} + \vec{\cal A} = B_{eff} x$ in this reference frame. Therefore, the mean field Dirac operator acquires the form
 	\begin{eqnarray}
 	\hat{Q}_{mean\, field} &=& i \gamma^0 \partial_t +\gamma^0  E_{eff} x + \mu\gamma^{\nu} u_{\nu} +  \gamma^k  \Bigl( i {\partial}_k + B_{eff} x\hat{y}_k \Bigr) - M  
 \end{eqnarray}
Here $E_{eff}$ and $B_{eff}$ differ from $E$ and $B$ by the factor $\frac{1}{1+2sp}$.
	The two dimensional Lorentz transformation representing a boost in the $y$-direction is given by 
	\begin{eqnarray}
		y &\to & \gamma (v) (y - v t ) \nonumber\\ 
		t &\to & \gamma (v)(t - v y) 
	\end{eqnarray}
	with relativistic $\gamma$-factor $\gamma (v)=(1-v^2)^{-\frac{1}{2}}$ under which the Chern - Simons gauge field, the external electromagnetic gauge field and the Dirac field transform as
	\begin{eqnarray}
		&{\cal A}_0 \rightarrow \gamma (v){\cal A}_0-\gamma (v)v\mathcal{A}_y, \,\,\,\,{\cal A}_y \rightarrow \gamma (v){\cal A}_y-\gamma (v)v\mathcal{A}_0,\\
		&A_0 \rightarrow \gamma (v)A_0-\gamma (v)vA_y, \,\,\,\, A_y \rightarrow \gamma (v)A_y-\gamma (v)vA_0,\\
		&\psi \rightarrow e^{\vec{v}\cdot \vec{\Sigma}}\psi = e^{\vec{v}\cdot \frac{\vec{\sigma}}{2}}\psi ,\,\,\,\,\vec{\Sigma}=\frac{1}{4}[\gamma^0,\vec{\gamma}], \,\,\,\, \gamma^0=\sigma^3,\,\,\,\,\gamma^1=i\sigma^2,\,\,\,\,\gamma^2=-i\sigma^1.
	\end{eqnarray}
	{As the fermion action is Lorentz invariant, a boost with velocity $v_y=\frac{E_{eff}}{B_{eff}}$ will only modify those contributions which result from picking a fixed reference frame (the one with respect to which the chemical potential $\mu$, electric field $\vec{E}$ and the magnetic field $\vec{B}$ are defined). We have $\vec{E}_{eff}=\vec{E}+\vec{\mathcal{E}}$ where $\vec{\mathcal{E}}$ is the electric field generated by the dynamical gauge field in analogy with the magnetic case.} {The result of the boost (with $|\vec{v}|\ll 1$) is
	\begin{eqnarray}
		\hat{Q}[{\cal A},\lambda] &=& i \gamma^0 \partial_t + \mu\gamma^{\nu}u_{\nu}^{\prime}+ \gamma^k  \Bigl( i {\partial}_k + B_{eff}x\hat{y}_k \Bigr) - M.  
\end{eqnarray}}

{{with $u^{\prime}_{\nu}=(1,0,-\frac{E_{eff}}{B_{eff}})$ to leading order. The constant term $\mu\gamma^{\nu}u_{\nu}^{\prime}$
{points out that chemical potential is defined in the laboratory reference frame (where electric field is nonzero) and not in the given reference frame (where there is no electric field). Gauge transformation $\psi \to e^{i (E/B) y }\psi$ removes the term $\gamma^2 (E/B)$ completely. Therefore, it does not affect physical results, and we arrive at the conclusion that in the refence frame without electric field the electrons are described by 
	\begin{eqnarray}
	\hat{Q}[{\cal A},\lambda] &=& i \gamma^0 \partial_t + \mu\gamma^{0}+ \gamma^k  \Bigl( i {\partial}_k + B_{eff}x\hat{y}_k \Bigr) - M.  
\end{eqnarray}
This is the system of electrons in the presence of external magnetic field $B_{eff}$. In this reference frame the electron liquid is at rest. Therefore, in the laboratory reference frame (with electric field) it moves with velocity $E/B$. }

\subsection{The Zubarev statistical operator}
\label{SectMod}

{In this section we will consider (in the framework of the relativistic extension of the theory by Lopez and Fradkin) the Zubarev statistical operator and its representation through the functional integral. At rest this
system is equivalent to Dirac fermions coupled to the statistical gauge field
with Chern - Simons action. At the same time, spatial components of the macroscopic four velocity
$u$ are to be given by $E/B$ being orthogonal to the electric field. The latter condition will appear
dynamically. The representation of the Zubarev statistical operator through a functional integral was considered specifically in \cite{selch2024effective}}.\par
{The Zubarev statistical operator is a relativistically covariant representation of the statistical operator $\hat{\rho}$ of the form} {
\begin{align}
\hat{\rho}=\frac{1}{Z}e^{-\int_{\Sigma}d^4z\beta n^{\mu}(T_{\mu\nu}u^{\nu}-\mu j_{\mu})}.
\label{statoperator}
\end{align}}
{The quantities that appear within the relativistic statistical operator are the inverse temperature $\beta$, the energy momentum tensor $T_{\mu\nu}$ with associated three velocity of macroscopic motion $u^{\mu}$ such that $u^{\mu}u_{\mu}=1$, the electric charge current $j^{\mu}$ with associated chemical potential $\mu$ as well as a spatial hypersurface $\Sigma$ with normal vector field $n^{\mu}$. It is assumed that space - time may be foliated by a one parameter family of hypersurfaces isomorphic to $\Sigma$. In the following we will assume that $\Sigma$ is a hyperplane in flat Minkowski space - time. This representation allows us to identify a Hamiltonian density in the following manner
\begin{align}
\mathcal{H}_{mm}=n^{\mu}(T_{\mu\nu}u^{\nu}-\mu j_{\mu}).
\end{align}
Time derivatives of fields within $\mathcal{H}_{mm}$ are subsequently eliminated by employing the Heisenberg equation of motion comprising the original Hamiltonian density $\mathcal{H}$ without macroscopic motion. By means of the path integral technique we may finally obtain a Lagrangian $\mathcal{L}$ comprising macroscopic motion.}
We consider relativistic electrons in the presence of a fluctuating statistical gauge field, and in the presence of the fluctuating electromagnetic field $\lambda$ around the fixed external electromagnetic field $A$. The partition function in Minkowski space - time is then given by (for details of the derivation see \cite{selch2024effective})
	\begin{equation}
		Z[{\mathfrak U}] = \int D\psi D\bar{\psi} D \lambda D{\cal A} e^{i S_{CS}[{\cal A}]  \mz{ + i\int d^3 x \mathcal{L}_f(\overline{\psi},\psi,A,\lambda,{\cal A},{\mathfrak U})+\int d^4 x \mathcal{L}_g(A,\lambda,{\cal A},{\mathfrak U}) ,\mathfrak{U})}}\label{ZU} 
	\end{equation}
	with 
	\begin{eqnarray}
		S_{CS} & = &  \frac{\theta}{4} \int d^3 z \epsilon^{ijk} {\cal A}_i {\cal F}_{jk} \nonumber\\ 
		{D} & = & {\partial} + i {A} \mz{+}i \lambda +i {\cal A}.
	\end{eqnarray}
	The procedure proposed in \cite{selch2024effective} had to be extended in order to include the statistical gauge field. However, the stress - energy tensor of the field with Chern - Simons action vanishes. As a result in the effective path integral representation of the Zubarev statistical operator the statistical gauge field $\cal A$ does not interact directly with the macroscopic velocity $\mathfrak{U}$ {(which will be related to the three velocity $u$ shortly)}. 
	The effective Lagrangian describing macroscopic motion of the Hall liquid together with the energy of the external electromagnetic field is then given by 
	{\begin{align}
	\mathcal{L}_f(\overline{\psi},\psi,A,\lambda,{\cal A},{\mathfrak U})=& \overline{\psi}(\gamma^{\mu}\frac i2\overset{\leftrightarrow}{D}_{\mu}-m)\psi 
	\nonumber\\	
	&+ { {\mathfrak U}_k\overline{\psi}\gamma^0(\frac{i}{8}[\gamma^{j},\gamma^k](\overset{\leftarrow}{D}_j+D_j) 		-\frac{i}{2} \overset{\leftrightarrow}{D^k})\psi}  \label{rez1_}\\
\mathcal{L}_g(A,\lambda,{\cal A},{\mathfrak U})=	&\frac{1}{2 e^2 }(\tilde{ E}_{i}+\tilde{\Lambda}_i)^2-\frac{1}{2e^2}({ B}+\Lambda)^2,\label{rez2_}
\end{align}
where
\begin{equation}
\tilde{E}_{i} = {E}_i -  \epsilon_{ij} {B} {\mathfrak U}^j, \quad \tilde{\Lambda}_{i} = {\Lambda}_i -  \epsilon_{ij} {\Lambda} {\mathfrak U}^j\label{Etilde}
\end{equation}
and 
$$
\Lambda_i = \partial_{[0} \lambda_{i]}\quad \Lambda  =  - \epsilon^{0ij} \partial_{[i} \lambda_{j]}/2 .
$$
}
{We have $\mathfrak{U}^{\mu}=\frac{u^{\mu}}{u^0}$ which implies the "gauge" ${\mathfrak U}_0 = 1$. Notice that in the non-relativistic limit $u^0\approx 1$ and therefore $\mathfrak{U}^{\mu}\approx u^{\mu}$.  It can be seen that the Lagrangian without macroscopic motion may be recovered in the limit $\mathfrak{U}^i\to 0$. The requirement that the expression for the Zubarev statistical operator given above is stationary (that is, the theory is in thermodynamical equilibrium) restricts the possible profiles of macroscopic velocity. Namely, 
motion along the straight line with constant velocity is admitted, together with rigid rotation and a special type of accelerated motion \cite{selch2024effective}. 	If we are interested in the simplest case of motion along a straight line, the gauge part of the effective action reduces to 
 \begin{align}
 	\mathcal{L}_g(A,\lambda,{\cal A},{\mathfrak U})=	&\frac{1}{2 e^2 }(\tilde{ E}^{i})^2-\frac{1}{2e^2}{ B}^2 + \frac{1}{2 e^2 }(\tilde{ \Lambda}^{i})^2-\frac{1}{2e^2}{ \Lambda}^2,\label{rez3_}
 \end{align}
\mz{The cross - term vanishes due to integration by parts when we consider the action given by the integral of the Lagrangian.}
It can be easily seen that the last term in Eq. (\ref{rez1_}) effectively results in the appearance of the addition to the electrostatic potential $A_0$ equal to $-\vec{\mathfrak U}\vec{A}$. \mz{In the general case an effective electrostatic potential  comprises also a term proportional to the divergence of $\mathfrak{U}$, but now it is absent since we consider the case of constant $\mathfrak{U}$.} For the chosen form of the vector potential this extra term in the electrostatic potential results in the replacement of the external electric field experienced by the fermions by $\tilde{E}^i$ given above in Eq. (\ref{Etilde}). It is well  known that the state with nonzero electric field cannot be stable due to pair production, and we come to the conclusion that only the choice of a specific value of macroscopic velocity leads to a stable theory and thermodynamical equilibrium. As expected, the corresponding value of velocity is     
\begin{equation}
	{\mathfrak U}^i  = \epsilon^{ij} E_j /B =\epsilon^{ij}\mathcal{E}_j/\mathcal{B}\label{UEB}
\end{equation}}

Notice also that according to the results of \cite{selch2024effective} the theory with the given constant macroscopic velocity appears as a result of the Lorentz boost with the same velocity applied to the original theory (with vanishing $\mathfrak U$). 
 
Thus the Zubarev statistical operator of the given theory with nonzero electric field describes a stable state of matter with $\tilde{E} = 0$ when the macroscopic velocity is given by Eq. (\ref{UEB}). This means that the system in the presence of constant electric field actually is not at rest {(relative to the laboratory frame)} but moves with this macroscopic velocity. If we apply a Lorentz boost to the system such that $\mathfrak{U}_i^{\prime}=0$ in the new reference frame, we come to its conventional description, now in a moving reference frame {(relative to the laboratory frame)}, where the real electric field vanishes. In this reference frame the number of electrons (or as well composite fermions) is governed by $B_{eff}$.
\ms{We illustrate the relation between the laboratory and Hall fluid reference frames schematically in Fig. (\ref{labvshallframe}).}
\begin{figure}
\begin{center}
\includegraphics[scale=0.1]{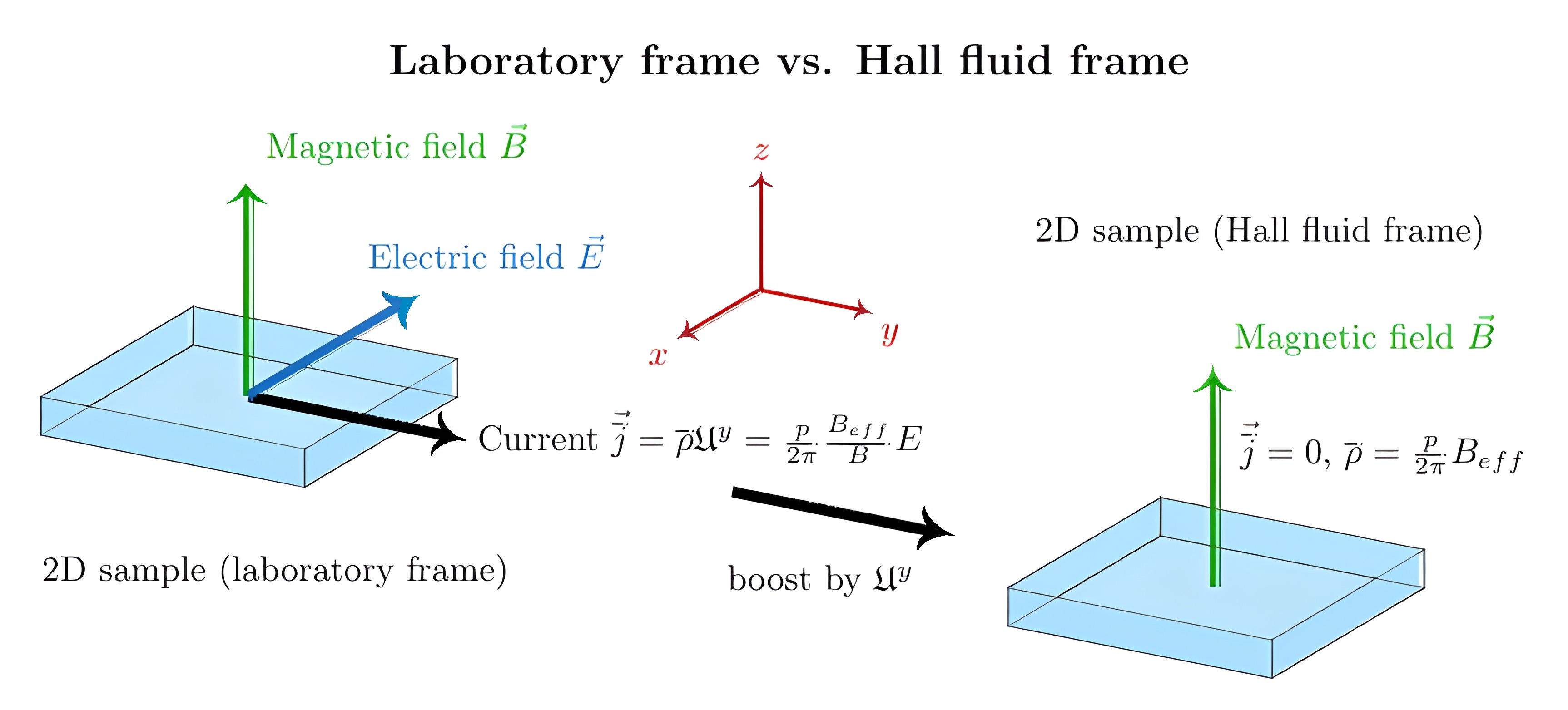}
\end{center}
\caption{Schematic illustration of the connection and properties of the laboratory and Hall fluid reference frames. While the laboratory reference frame features a nonzero Hall current and electric field, these vanish in the Hall fluid frame.}
\label{labvshallframe}
\end{figure}
 
{Interestingly, the second line of the macroscopic motion Lagrangian in Eq. (\ref{rez1_}) provides the dipole and vorticity terms  as advertised in \cite{nguyen2018particle} according to  
\begin{align}
\epsilon^{\mu\nu\lambda}\frac{F_{\nu\lambda}}{2B}\overline{\psi}\frac{i}{2}\overset{\leftrightarrow}{D}_{\mu}\psi =\overline{\psi}\gamma^0\frac i2\overset{\leftrightarrow}{D}_0\psi -{\mathfrak U}_k\overline{\psi}\gamma^0\frac{i}{2} \overset{\leftrightarrow}{D^k}\psi
\end{align}
as well as
\begin{align}
\frac{1}{2}\frac{\nabla_iE^i}{2B}\overline{\psi}\psi \cong{\mathfrak U}_k\overline{\psi}\gamma^0\frac{i}{8}[\gamma^{j},\gamma^k](\overset{\leftarrow}{D}_j+D_j)\psi .
\end{align}}

where $\cong$ means equality up to partial integration in the integrated expressions.

\mz{Importantly, in the above considerations of this section we did not rely on the mean field approximation, and considered the theory precisely. Thus the obtained results allow us to reduce the calculation of Hall conductivity to the density of electrons $\rho$ in the moving reference frame (with velocity of Eq. (\ref{UEB})). All these electrons in the laboratory reference frame move with the given velocity, and we will arrive at the value of QHE conductivity $\rho/B$. In the next sections we will consider $\rho$ with all perturbative corrections taken into account while in the rest of this section we restrict ourselves to its mean field value.}   

 At the level of the mean field approximation the electron liquid moving with velocity $\mathfrak U$ given in Eq. (\ref{UEB}) looks like a collection of $p$ occupied Landau levels of the effective magnetic field $B_{eff} = B/(1+2sp)$. The density of electrons (or as well composite fermions) is \mz{$\bar{\rho}=p\frac{B_{eff}}{2\pi}$}. All the corresponding electrons at zero temperature
move with velocity $E/B$, and we arrive at the  electric current density 
{\begin{equation}
j_y = \overline{\rho}\mathfrak{U}_y=p\frac{B_{eff}}{2\pi}\frac{E}{B}
\end{equation}}
and the QHE conductivity 
\begin{equation}
\sigma_{xy} = \frac{1}{2\pi}\frac{p}{1+2sp}.
\end{equation}
Thus already at the level of the mean field approximation we again come to the original value of the fractional QHE in the given system, obtained in the previous sections using different methods.

\subsection{Interactions due to fluctuations of $\lambda$ and $\cal A$ above their mean values}

According to the results of the previous section in order to calculate the QHE conductivity it is enough to calculate density of electrons $\rho$ in the reference frame moving with velocity of Eq. (\ref{UEB}). In turn, this value of density is given by $\rho = i \frac{d Z}{V\, Z d \mu} $, where $V$ is the overall volume of the system. Therefore, in order to calculate the interaction corrections to the QHE conductivity it is enough to consider the partition function itself. Thus we represent the statistical gauge field as a sum of its mean value ${\cal A}_0$ and fluctuation above it ${\cal A}'$. For the field $\lambda$ the mean value is zero, and therefore fluctuation above it is denoted by the same letter.  The partition function (in the moving reference frame, where there is only the external magnetic field, and no electric field) may then be represented as follows
\begin{equation}
	Z = \int D\psi D\bar{\psi} D \lambda D{\cal A}'  e^{i S_{CS}[{\cal A}_0+{\cal A}'] + iS_g[\lambda]  + i\int d^3 x \bar{\psi} (\hat{Q}[{\cal A}+{\cal A}',\lambda]) \psi} 
\end{equation}
with 
\begin{eqnarray}
	S_g & = &  - \frac{1}{4 e^2}\mz{\int d^4z} (\partial_{[\mu} (\lambda)_{\nu]})^2 + \mz{\int d^4z}L_{gauge\, fixing}[\lambda]\nonumber\\ 
	S_{CS}[{\cal A}] & = &  \frac{\theta}{4} \int d^3 z \epsilon^{ijk} {\cal A}_i {\cal F}_{jk} \nonumber\\ 
	{D} & = & {\partial} + i {A}+ i{\cal A}_0  + i {\cal A}' +i\lambda   
\end{eqnarray}}

The Dirac operator is
\begin{eqnarray}
\hat{Q}[{\cal A},\lambda] &=& i \gamma^0 \partial_t -\gamma^0 {\cal A}'_t + \mu\gamma^{\nu} u_{\nu} - \gamma^{\mu} \lambda_\mu(z)+ \gamma^k  \Bigl( i {\partial}_k + B_{eff}x\hat{y}_k - {\cal A}'_k\Bigr) - M  \label{QE0}
\end{eqnarray}
where $u = (1,0,-E/B)$ in the {\it given} reference frame {(the moving frame with zero $E$)}.

The Chern - Simons term in the action may be represented as
\begin{equation}
	S_{CS}[{\cal A}_0+{\cal A}']  = S_{CS}[{\cal A}_0]+S_{CS}[{\cal A}'] +  {\theta} \int d^3 z {\cal A}'_t {B}\frac{2sp}{1+2sp}
\end{equation}

In Appendix \ref{AppCSfluct} we demonstrate that there is no effect on the partition function of the interactions due to exchange by fluctuations of statistical gauge field (above its mean value). Therefore, in order to calculate the electron density the statistical gauge field may be replaced safely by its mean value. The only interactions that remain relevant are those due to exchange by quanta of $\lambda$.

\section{Non - renormalization of fractional QHE conductivity by interactions}
\label{SectInt}
\subsection{One loop electron self energy}

\label{SectSigma}

Thus in order to go beyond the mean field approximation we may consider the electron liquid in the moving reference frame and calculate the density of electrons including the effect of interactions (due to the fluctuations of $\lambda$ above its mean field value $0$). As we discussed above, there is no effect of the fluctuations of the statistical gauge field on the QHE conductivity. Strictly speaking, outside the mean field approximation the description with the aid of Landau levels fails. 

In the leading order we neglect interactions with fluctuations of the fields above their mean field values $\lambda_0 = 0, {\cal A}_{\mu} = (-\frac{1}{1+2sp}Ex,0,\frac{1}{1+2sp}Bx)$. Above it has been proven that fluctuations of statistical gauge field do not affect the expression for the electron density, therefore we take into account only the fluctuations of electromagnetic field $\lambda$. We then calculate the electric charge density as 
\begin{equation}
	\rho = \mzc{-}i \mathrm{Tr}\, \int \frac{d^3 p}{(2\pi)^3} {G}_W(x,p)
	\partial_{p^0} Q_W(x,p)\label{rho0}
\end{equation}
where the Dirac operator $\hat{Q}$ is given by 
\begin{eqnarray}
	\hat{Q}=\hat{Q}_{mean\, field} &=& i \gamma^0 \partial_t  + \mu\gamma^{\nu} u_{\nu} - \gamma^{\mu}\lambda_\mu(z)+ \gamma^k  \Bigl( i {\partial}_k + \frac{1}{1+2sp}Bx\hat{y}_k\Bigr) - M  +i0
\end{eqnarray}
while $\hat{G}$ is its inverse (with the poles in $p_0$ shifted from the real axis is such a way that the resulting Green function represents the T - ordered, i.e. Feynman propagator). The subscript $W$ points to the use of the Wigner - transformed expressions.

We perform Wick rotation, and the calculations are performed in
Euclidean space - time, where there is no ambiguity in the ways to avoid poles during integration over $p_0$.
In the next to leading order for the calculation of the density (and thus the conductivity) we
	take into account the one - loop electron self energy $\hat \Sigma$ due to
	exchange by quanta of the gauge field $\lambda$.  Then the renormalized Dirac operator is
	given by  
	\begin{equation}
		\hat{\mathbf{Q}} = \hat{{Q}}_{eff} - \hat{\Sigma},
	\end{equation}
	where $\hat{{Q}}_{eff}$ is Dirac operator in the presence of the
	effective magnetic field $B_{eff}$.

In the present section we consider the rainbow approximation (when the self
	energy operator of the gauge boson is neglected, as well as the
	interaction corrections to the vertex with emission of  gauge field
	quantum). 

Consider the self energy in
	configuration space at lowest order in perturbation theory
	\mz{\begin{align}
		\Sigma(x,y)=&D^{ij}(x,y)\gamma^iG(x,y)\gamma^j.  \nonumber
\end{align}}
Here $D^{ij}$ is the propagator of the gauge field $\lambda_{\mu}$, the photon.

The Weyl symbol of the self-energy operator finally acquires the form  
\begin{align}
	\Sigma_W(x,p)=&\int d^3r\Sigma(x+\frac{r}{2},x-\frac{r}{2})e^{-ipr} 
	\nonumber \\
	=&\int \frac{d^3q}{(2\pi )^3} \mz{D_W^{ij}(x,q)\gamma^iG_W(x,p-q)%
	\gamma^j.}
\end{align}

\subsection{$\rho$ and $\rho^\prime$}

{As above, we solve the Groenewold equation 
	\begin{equation}
		\mathbf{G}_W \star \mathbf{Q}_W = 1
	\end{equation}
	iteratively by a derivative expansion. The leading terms in this expansion give the average electron density (after Wick rotation to three dimensional space - time of Euclidean signature) 
	\begin{eqnarray}
		\rho & = & \mzc{-i} {\frac{T}{ (2\pi)^3 \mathcal{S}}} \, \int \,\mathrm{Tr}\,
		d^3p \, d^3x\,  \Big[ \mathbf{G}_W(p,x)\star \frac{%
			\partial (Q_{eff})_W(p,x)}{\partial p_3} \Big]\,.  \label{calM3}
	\end{eqnarray}
	Here $S$ is the area of the system (supposed to be large), $T$ is temperature (supposed to be small), $\hat{\mathbf{Q}}_{} = \hat{Q}_{eff} - \hat{\Sigma}$ is the
	renormalized Dirac operator, while $\hat{Q}_{eff}$ is the free Dirac operator in
	the presence of external magnetic field $B_{eff}$. The Green function $\hat{%
		\mathbf{G}}_{}$ obeys 
	\begin{equation}
		\mathbf{G}_{W} \star \mathbf{Q}_{ W} = 1
	\end{equation}
}

{Let us define 
	\begin{eqnarray}
		\rho^{\prime }& = & \mzc{-i}{\frac{T}{ (2\pi)^3 \mathcal{S}}} \, \int \,%
		\mathrm{Tr}\, d^3p \, d^3x\, \, \Big[ \mathbf{G}_W(p,x)\star  \frac{%
			\partial \mathbf{Q}_W(p,x)}{\partial p_3} \Big]\,.
		\label{calM5}
	\end{eqnarray}
It may be proven easily that it is a topological invariant:  variation of $\bf Q$ does not cause a variation of $\rho'$. Therefore, its value remains equal to the bare density of electrons $\frac{p}{1+2sp}\frac{B}{2\pi}$ corresponding to the number
	of occupied effective Landau levels.}

%\label{SectGed}

{For a squared sample of linear size $L$ the difference between $\rho$ and $\rho'$  is given by 
	\begin{equation}
		\langle\Delta \rho\rangle = \mzc{-i}\frac{1}{L^2}\mathrm{Tr}\, \int d^2x\int \frac{%
			d^3 p}{(2\pi)^3} \mathbf{G}_W(x,p) \partial_{p^3} \Sigma_W(x,p) = \mzc{-i}\frac{1}{%
			L^2}\mathrm{Tr}\, \int d^2x\int \frac{d^3 p}{(2\pi)^3} \mathbf{G}_W(x,p)
		\star\partial_{p^3} \Sigma_W(x,p)
		\label{electriccurrentdifference}
	\end{equation}
	Here we should use the expression for $\Sigma_W$ obtained in Sect. \ref%
	{SectSigma} with the propagator of the  gauge boson ($D_{\mu\nu}(p)$). The renormalized propagator (with polarization operator taken into account) becomes dependent on both incoming and outgoing momenta of the photon. 
	It's Wigner transformation is denoted by $\mathbf{D}^{\nu\mu}_W(z,p)$. It obeys the symmetry property
	\begin{equation}
		\mathbf{D}^{\nu\mu}_W(z,-p) = \mathbf{D}^{\mu\nu}_W(z,p).
	\end{equation}
 \ms{We do not provide explicit expressions for the bare and renormalized photon propagators $D^{\mu\nu}$ and $\mathbf{D}^{\mu\nu}$ intentionally. The reason is that we would then need to make the gauge fixing term in the Lagrangian explicit. Our arguments work by just exploiting the symmetry properties of the bare and renormalized photon propagators which are gauge independent. This may be checked in a straightforward way for the bare propagator. The renormalized photon propagator retains the symmetries of its bare form, as may be concluded with the help of Sect. 6.4.4 of \cite{Zhang2022} adapted to vector boson as interaction mediators.}
	
	\subsection{ $\Delta \rho$ at first loop order}
	
	\label{deltarho}
	
	In the "relativistic" version of the model by Lopez and Fradkin, which replaces the non
	- relativistic Hamiltonian with the Dirac Hamiltonian, we substitute bare vertex $\gamma^\mu$ for photon emission 
		and therefore 
	\begin{eqnarray}
		\langle\Delta \rho \rangle &=& \mzc{-i}\frac{1}{L^2}\mathrm{Tr}\, \int d^2x\int \frac{%
			d^3 p}{(2\pi)^3}\frac{d^3 k}{(2\pi)^3} \mathbf{G}_W(x,p) 
		\gamma^\mu \partial_{p^3} \mathbf{G}^{(0)}_W(x,p-k)  \gamma^\nu 
		\mathbf{D}_W^{\mu\nu}(x,k)  \nonumber \\
		&=& \mzc{i}\frac{1}{L^2}\mathrm{Tr}\, \int d^2x\int \frac{d^3 p}{(2\pi)^3}\frac{%
			d^3 k}{(2\pi)^3} \partial_{p^3}\mathbf{G}_W(x,p) \gamma^\mu 
		\mathbf{G}^{(0)}_W(x,p-k)  \gamma^\nu \mathbf{D}_W^{\mu\nu}(x,k) 
		\nonumber \\
		&=& \mzc{i}\frac{1}{L^2}\mathrm{Tr}\, \int d^2x\int \frac{d^3 p^{\prime }}{(2\pi)^3%
		}\frac{d^3 k^{\prime }}{(2\pi)^3} \partial_{{p^{\prime }}^3}\mathbf{G}%
		_W(x,p^{\prime }-k^{\prime})\gamma^\mu \mathbf{G}^{(0)}_W(x,p^{\prime})
		\gamma^\nu \mathbf{D}_W^{\mu\nu}(x,-k^{\prime })  \nonumber \\
		&=& \mzc{i}\frac{1}{L^2}\mathrm{Tr}\, \int d^2x\int \frac{d^3 p^{\prime }}{(2\pi)^3%
		}\frac{d^3 k^{\prime }}{(2\pi)^3} \mathbf{G}^{(0)}_W(x,p^{\prime})
		\gamma^\nu \partial_{{p^{\prime }}^3}\mathbf{G}_W(x,p^{\prime }-k^{\prime})
		\gamma^\mu \mathbf{D}_W^{\mu\nu}(x,-k^{\prime })  \nonumber \\
		&=& \mzc{i}\frac{1}{L^2}\mathrm{Tr}\, \int d^2x\int \frac{d^3 p^{\prime }}{(2\pi)^3%
		}\frac{d^3 k^{\prime }}{(2\pi)^3} \mathbf{G}^{(0)}_W(x,p^{\prime})
		\gamma^\nu \partial_{{p^{\prime }}^3}\mathbf{G}_W(x,p^{\prime }-k^{\prime})
		\gamma^\mu \mathbf{D}_W^{\nu\mu}(x,k^{\prime })  \label{currentdifference}
	\end{eqnarray}
	We made a change of variables $-k = k^{\prime }$, $p-k = p^{\prime }$, and perform integration by parts with respect to variable $p^3$. This can be done because expression standing inside the integral approaches zero at $p^3 \to \pm \infty$.
	One can see that $\langle \Delta \rho \rangle = -\langle \Delta \rho \rangle =
	0$. The logic of the above derivation may be extended to the higher orders of perturbation theory along the lines of \cite{Zhang2022}, which will be discussed below.
	
	\subsection{ $\Delta \rho$ at higher orders}
	
	{\ms{In the previous section we presented the proof that
		$\Delta \rho = 0$ if we take the exchange by quanta of the gauge field
		in the rainbow approximation (note that all propagators where fully interacting ones). A general proof to all orders of
		perturbation theory (not restricted to the rainbow approximation) in the interactions due to 
		$\lambda$ is obtained following the methodology of \cite{Zhang2022} (see section VI. of that work).} The fermions in \cite{Zhang2022} were coupled by a Yukawa term to a scalar field $\phi$. The proof relates the perturbatively renormalized current at each loop order to a sum of the so-called progenitor diagrams (Feynman diagrams without external current insertion) supplemented by a total derivative operator inserted into the momentum integral of a fermion loop identical to the loop momentum (which therefore vanishes identically). This equality has been achieved in such a way as to cancel the symmetry factors present in the sum of progenitor diagrams. Symmetry factors are absent after current insertions into the progenitor diagrams turning them into contributions of the electromagnetic current. The bosonic propagators considered in \cite{Zhang2022} are undirected. Our present case differs from the case considered in \cite{Zhang2022} in the following three aspects:
		\begin{enumerate}
			\item In \cite{Zhang2022}  the interaction corrections to the electric current  are considered. We consider the corrections to charge density of Eq. (\ref{calM3}). The difference between the two expressions is that in case of \cite{Zhang2022} the derivative $\partial_{p_k}$ of the non - interacting Dirac operator is with respect to the spatial component $p_k$ of momentum while in our case it is with respect to Matsubara frequency $p^3$, and is multiplied by $i$.  
			
		\item
		
		In \cite{Zhang2022} the $2D$ system on torus is considered. The torus is divided into the two equal pieces, the absolute value of external electric field is constant, while its directions are opposite within the two mentioned pieces  of the torus. In our case we consider the system in the reference frame, where electric field vanishes at all. 
		
		\item 
		In \cite{Zhang2022} the interactions due to exchange by scalar electric potential $\phi$ are considered with the interaction vertex $\overline{\psi}\psi \phi$.
		In our case the electromagnetic field $\lambda_\mu$ is coupled to fermions via $j^{\mu} \lambda_{\mu}=\overline{\psi}\gamma^{\mu}\psi\lambda_{\mu}$.

		\end{enumerate}
	
		  The proof of \cite{Zhang2022} may be  trivially extended to our case as the Feynman diagrams remain unmodified. In fact, all the content of Sections VI.4.1, VI.4.2, VI.4.3, VI.4.4 of \cite{Zhang2022} (except for Eq. (100)) is valid for our case with the replacement of $\partial_{p_k}$ ($k = 1,2$) by $i \partial_{p^3}$, and replacement of scalar propagator $D$ by photon propagator $D^{\mu\nu}$, and the interaction vertex replaced by $\gamma^\mu$. 
		  In particular, the content of Sect. VI.4.1 of \cite{Zhang2022} has been repeated above in Sect. \ref{deltarho} of the present paper.

		   The integration by parts with respect to $p_k$ is an essential part of the proof given in \cite{Zhang2022}, it may be performed due to the periodic boundary conditions resulted from the torus geometry of the sample. In our case we perform integration by parts with respect to $p^3$. It is performed due to the fact that the electron propagator tends to zero at $p_3 \to \infty$, which makes vanishing the expressions standing inside the integrals at infininte values of $p_3$. Notice that the Feynman rules for the diagrams written in terms of the Wigner transformed propagators is an essential tool used for the treatment of the higher orders of perturbation theory in this proof. This formalism has been described in details in Sect. 5 of \cite{Zhang2022}. {\it We repeat this description here with the corresponding changes in notations in Appendix \ref{sectbubble}.}}\par
	   
	    {\it Moreover, we also give in Appendix \ref{app3} the detailed consideration itself of one - loop, two - loop and higher orders corrections to the electron density.}

We come to the conclusion that the density of electrons in the moving reference frame (where electric field vanishes) remains unaffected by interactions both due to the exchange by virtual photons $\lambda$ and due to the statistical gauge field (for the latter see Appendix \ref{AppCSfluct}).
As a result we arrive at the value of QHE current $\frac{E}{B}\, \times \frac{B}{2\pi}\times \frac{p}{1+2sp}$, which gives the QHE conductivity $\sigma_{QHE} = \frac{1}{2\pi}\,\times \frac{p}{1+2sp}$.
	
This is the proof that the fractional QHE conductivity is not influenced by interactions as long as we consider the latter perturbatively. \ms{Notice that the non-renormalization proof merely considers the dynamics of electrons and not that of the lattice of ions which serve as a passive background in our proof. At the mean field level we may incorporate the effects of a positive charge background by a vacuum mixing mediated by $\lambda$ in the non-relativistic and by $\lambda_{\mu}$ in the relativistic case. This will not modify the logic of the arguments for Hall conductivity non-renormalization. The explicit inclusion of an external magnetic field modifies the Feynman diagrams in an identical way. Including dynamical background effects beyond the mean field level as just indicated for the lattice of ions or otherwise an exclusive treatment of the electrons within the Keldysh technique will invalidate the proof for conductors due to dissipation effects arising from interactions with the background ion lattice. The reduction of the discussion to electrons or composite fermions in diagrammatic perturbation theory is sufficient for the insulating quantum Hall phases where only the Hall response transverse to the applied electric field survives, while the longitudinal component vanishes identically. It is worth mentioning that there may exist analogies between the approach  presented here to that of \cite{gutman2011non,skorobagatko2020self}. Finally note that our considerations are not material specific and valid rather generally.}

\section{Conclusions}

\label{SectConcl}

In the present paper we reconsidered the theory of the fractional QHE by Lopez and Fradkin. First of all, we emphasize that the seminal result announced in \cite{Fradkin1991chern} (in the units of $e^2/\hbar$) for the principal Jain series
\begin{equation}
\sigma_{xy} = \frac{1}{2\pi}\times \frac{p}{1+2sp} \label{GJS}
\end{equation}
is valid already at the level of mean field theory. This occurs because the external electric field is screened by the statistical gauge field in a way similar to the external magnetic field. This reasoning can also be read off  from Sect. 3.1. of \cite{simon1998chern}. Roughly we can explain this as follows: in the reference frame moving with velocity $E/B$ there is only the external magnetic field, and there is no external electric field. When we perform a Lorentz boost to the laboratory reference (or rest) frame (with electric field present), the screening magnetic field gives rise to a screening electric field due to the transformation of the electromagnetic field. \mz{Notice that the screening of electric field is also intimately related to the well - known appearance of effective fractional charge $\frac{e}{1+2sp}$ of electron quasiparticles.}

Next, we explore the relativistic extension of the theory (that differs from that of \cite{son2018dirac,nguyen2018particle}). 
We propose to look at the given theory from an alternative point of view based on the Zubarev statistical operator, and its representation in the form of the functional integral. This way we describe the electron liquid with QHE by macroscopic motion of the field system. The corresponding velocity is given by the ratio $u = E/B$, which is the only case giving a stable effective theory. The electric current is then obtained as the flow of electrons existing in the moving reference frame (where the electron liquid is at rest as a whole). We again come to the same expression for the fractional QHE conductivity:
$$
\sigma_{xy} = p \frac{u}{E} \frac{B_{eff}}{2\pi} = \frac{p}{B} \frac{B_{eff}}{2\pi} =  \frac{1}{2\pi}\frac{p}{1+2sp}
$$ 
In order to prove that the fractional QHE conductivity is not influenced by interaction corrections we follow the lines of \cite{Zhang2022}.
But our present consideration remains more simple: we only need to consider the corrections to the expression for the electron density in the moving reference frame, where there is no electric field. The output is that the conventional expression for the fractional QHE conductivity is not renormalized by interactions as long as the latter are taken into account perturbatively, and as long as the value of the chemical potential provides that the two - point Green function does not have poles (in Euclidean space - time).

Notice that following \cite{Fradkin1991chern} we considered the simplest version of the theory related to the LLL and describing the principal Jain sequence of Eq. (\ref{GJS}). We expect that the extension of our consideration to a more involved theory describing the other sequences \cite{lopez2004fermion,chang2004microscopic}, might be straightforward. However, the consideration of this issue remains out of the scope of the present paper. 

It is also worth mentioning that our consideration (as well as the consideration of the original paper by Lopez and Fradkin \cite{Fradkin1991chern}) was limited to the simplest Hamiltonian of free electron (quadratic in momentum), and the Dirac operator of the form of Eq. (\ref{ZLF0}). It would be important to extend our consideration to the theory with the one - particle Hamiltonian depending on momentum arbitrarily. It is natural to  suppose that this might be done in the same spirit even without any relativistic extension applying a kind of Galilean transformation instead of the Lorentz transformation. A non - relativistic version of the Zubarev statistical operator formalism should then be considered. We expect the appearance of certain difficulties on this way and intend to come to the consideration of these issues in forthcoming publications. 

\vspace{0.5cm}

\centerline{\bf Appendixes}

\appendix
\numberwithin{equation}{section}

\section{Notations of the calculus of differential forms }

\label{SectDiff}

 If $\Sigma$ is a $d$ - dimensional hypersurface in $D$ - dimensional space - time, while $\omega$ is a $d$ - form, then we denote $\langle \omega, \Sigma \rangle = \int_\Sigma \omega$. In addition, for each covariant tensor field $T$ of dimension $d$ and a $d$ - form $\omega$ we denote the action of $\omega$ on $T$ as $\langle \omega, T \rangle = \int \omega \wedge *T$, where integration is over the whole space - time. Therefore, by $\Sigma$ we denote both a hypersurface and  vector field $\Sigma(x) = \int_\Sigma dz^{\mu_1} \wedge ... dz^{\mu_{d}} \delta^{(D)}(x-z)\partial_{\mu_1}...\partial_{\mu_d}$. By $*$ we denote the Hodge duality operation that acts as follows. If $T \equiv T^{\mu_{D+1-d}...\mu_D}\partial_{\mu_{D+1-d}}...\partial_{\mu_D}$ then $*T = (*T)_{\mu_{1}...\mu_{D-d}}dx^{1}\wedge...\wedge dx^{D-d}= \frac{1}{d!} \epsilon_{\mu_1...\mu_D} T^{\mu_{D+1-d}...\mu_D}dx^{1}\wedge...\wedge dx^{D-d}$. The Hodge dual also may be defined for a $d$ - form: In components we have $*\eta = \frac{1}{d!} \epsilon^{\mu_1 ... \mu_D} \eta_{\mu_1 ... \mu_{d}} \partial_{\mu_{d+1}}...\partial_D$. One can check that $** = (-1)^{D-1}$ for space - time with Minkowski signature $(+,-,...,-)$, where $\epsilon_{01...(D-1)} = (-1)^{D-1}  \epsilon^{01...(D-1)} = (-1)^{D-1}$. For a $D-d$ form $\eta$ and $d$ - form $\omega$ we have $\int \eta \wedge **\omega = \langle \eta, *\omega\rangle = (-1)^{(D-d)d}\langle  \omega, *\eta \rangle = (-1)^{(D-d)d}\int  \omega \wedge **\eta$. We use the exterior derivative $d$ of a differential form defined in a usual way $d \omega_{i...k}dx^i \wedge... dx^k = \partial_j \omega_{i...k} dx^j \wedge dx^i\wedge... dx^k$. The dual operation (codifferential) acting on a covariant $d$ - dimensional antisymmetric tensor $T$ is defined as $\delta = (-1)^d(*)^{-1}d*$. One can check that it obeys $\langle d \omega, T\rangle = \langle \omega, \delta T\rangle $. Lowering and rising indices results in the correspondence between $d$ - form $\omega$ and $d$ - dimensional tensor: $
\urcorner \omega$, and vice versa - the $d$ - dimensional anti - symmetric tensor $T$ determines the $d$ - form $\lrcorner
T$. This allows to extend the action of operator $d$ to the antisymmetric tensors as $d T \equiv \urcorner d \lrcorner T$, and to extend the action of operator $\delta$ to the differential forms as $\delta \omega \equiv \lrcorner \delta \urcorner \omega$. Obviousely, $d^2 = \delta^2 = 0$. The space - time Laplace operator (d'Alembert operator) acting on a differential form is then defined as $\square = (\delta d + d \delta) $. Finally, we define for a differential form $\omega$ and antisymmetric vector field $T$: $\langle T, \omega\rangle \equiv \langle \lrcorner T, \urcorner \omega \rangle$ so that we can exchange both inside the brackets.

\section{Duality transformation applied to the statistical gauge field}

\label{AppDual}

In the present section we will use the calculus of differential forms in flat space - time  (with the real - valued coefficients). The corresponding notations are summarized in Appendix \ref{SectDiff}.

Here we prove using duality transformation the statement by Lopez and Fradkin \cite{Fradkin1991chern} that the standard model of electrons interacting by Coulomb forces is equivalent to their model 
	defined by the partition function of Eq. (\ref{Z1}). The former theory has the partition function
\begin{equation}
	Z = \int D\psi D\bar{\psi} D \lambda  e^{
		iS[\lambda] + i\int d^3 x \bar{\psi} (\hat{Q}[\mathcal{A},\lambda]) \psi}
	\label{Z0}
\end{equation}
with  
\begin{eqnarray}
	\hat{Q}[\lambda] &=&  i\partial_0 - A_0(z) +
	\mu - \lambda(z)+ \frac{1}{2M} \vec{D}^2  \nonumber \\
	S & = & \mz{\frac{1}{2} \int d^3 z d^3 \lambda(z) V^{-1}(z,z^{\prime })
		\lambda(z)}    \nonumber \\
	\vec{D} & = & \vec{\partial} - i \vec{A} \label{ZLF0}
\end{eqnarray}
The given partition function may be rewritten as an integral over the worldlines of the electrons in the following way
\begin{equation}
	Z = \int D \lambda \sum_{n}\Pi_{i = 1... n} (-1)^n Dz^{(i)}  e^{
		iS[\lambda] + i\sum_{j = 1...n}\int_{s = 0}^{1} ( A_\mu dz^{(j)\mu}(s) +  \lambda dz^{(j)0}+{\cal L}[z^{(j)}(s)])ds + iS_{int}[z(s)]}
	\label{Z0p}
\end{equation}
Here the sum is over the number $n$ of fermion loops. By ${\cal L}[z^{(j)}(s)]$ we denote the local one - particle Lagrangian depending on the worldline of the given loop, while $S_{int}[z]$ is an interaction term that accounts for the possible intersections (and self - intersections) of the loops.  One of the possible ways to derive representation of Eq. (\ref{Z0p}) is to use lattice regularization, and the standard technique of lattice quantum field theory that reduces the integral over the fermionic fields to the sum over the particle trajectories. This is the so - called "polymer representation" (see \cite{Montvay} and references therein). The continuous limit of the resulting expression is to be considered then. We will not be interested in the particular form of  ${\cal L}[z^{(j)}(s)]$ and $S_{int}[z]$. For us it is important that the only way how the fields $A$ and $\lambda$ enter the consideration is the one pointed out in Eq. (\ref{Z0p}). The similar "polymer representation" for the theory by Lopez and Fradkin of Eq. (\ref{Z1}) gives expression for partition function
\begin{eqnarray}
	Z &=& \int D \lambda \sum_{n}\Pi_{i = 1... n} (-1)^n Dz^{(i)}  e^{
		iS[\lambda] + i\sum_{j = 1...n}\int_{s = 0}^{1} ( A_\mu dz^{(j)\mu}(s) +  \lambda dz^{(j)0}+{\cal L}[z^{(j)}(s)]ds) + iS_{int}[z(s)]}\nonumber\\ &&\int D{\cal A} \,e^{i S_{CS}[{\cal A}] + i\sum_{j = 1...n}\int_{s = 0}^{1} {\cal A}_\mu dz^{(j)\mu}(s)}
	\label{Z1p}
\end{eqnarray}
We will be interested in the second row of this expression, and will show that it is equal identically to a constant. We denote this quantity by $Z[J]$, where $J$ is the given configuration of the worldlines composed of $n$ loops given by functions $z(s), s \in [0,1)$.

According to the above notations we denote by $J$ also the vector field corresponding to the electron worldline  
$$
J^\mu(x) = \int_J dz^\mu \delta(x-z) 
$$ 
The scalar product of this field and the statistical gauge field ${\cal A}$ is $\langle {\cal A}, J\rangle = \int_J {\cal A}dz$. The duality transformation acts as 
$$
[*{\cal A}]^{\mu\nu} =  \epsilon^{\rho\mu\nu} {\cal A}_\rho
$$
and for the field strength ${\cal F}_{\mu\nu} \equiv [d{\cal A}]_{\mu\nu}= \partial_\mu {\cal A}_\nu - \partial_\nu {\cal A}_\mu$:
$$
[*{\cal F}]^{\mu} = \frac{1}{2} \epsilon^{\rho\nu\mu} {\cal F}_{\rho\nu}
$$
 The Chern - Simons term then may be written as
$\frac{\theta}{2} \langle {\cal A},* {\cal F}\rangle $
  
Therefore, we rewrite
\begin{eqnarray}
	Z[{J}] &=& \int D{\cal A} \,e^{i \frac{\theta}{2} \langle {\cal A}, *{\cal F}\rangle + i\langle {\cal A}, J\rangle}\nonumber\\ &=& \int D{\cal A} D\Lambda  \,e^{i \frac{\theta}{2} \langle \square^{-1}\delta{\cal F}+d\alpha, *{\cal F}\rangle + i\langle {\cal A}, J\rangle}\delta(\Lambda - {\cal F})
	\nonumber\\ &=& \int D{\cal A} D\Lambda \,e^{i \frac{\theta}{2} \langle {\cal F}, \square^{-1} d*{\cal F}\rangle + i\langle {\cal A}, J\rangle}\delta(\Lambda - d{\cal A})
	\nonumber\\ &=& \int D{\cal A} D\Lambda D {\cal G}\,e^{i \frac{\theta}{2} \langle \Lambda, \square^{-1} d*{\Lambda}\rangle + i\langle {\cal A}, J\rangle + i \langle (\Lambda - d{\cal A}), {\cal G}\rangle}
		\nonumber\\ &=& \int D{\cal A} D\Lambda D {\cal G}\,e^{i \frac{\theta}{2} \langle (\Lambda+d*{\cal G}/\theta), \square^{-1} d*{(\Lambda+d*{\cal G}/\theta)}\rangle - \frac{i}{2\theta}\langle {\cal G}, *\delta  {\cal G}\rangle+ i\langle {\cal A}, J\rangle - i \langle   d{\cal A}, {\cal G}\rangle}
	\nonumber\\ &=& const \int   D {\cal G}\,e^{-i \langle {\cal G},*\delta  {\cal G}\rangle \frac{1}{2\theta}  }\delta(\delta {\cal G} - J)\nonumber\\ &=& {const}\,\int   D \Phi\,e^{-i \langle *d\Phi + \Sigma[J], *\delta(* d \Phi + \Sigma[J]) \rangle \frac{1}{2\theta}  }\label{Phi}
\end{eqnarray}
Here in the second row we use the representation for $\cal A$ through $\cal F$ and a gauge transformation $d\alpha$: 
$$
{\cal A} = \square^{-1} \delta {\cal F} + d\alpha
$$
In transition to the third row we use $\langle d \alpha, * {\cal F}\rangle = \langle \alpha, \delta * {\cal F} \rangle = \langle \alpha, * d {\cal F} \rangle = 0$. In the $5$-th line we use that $d\Lambda = 0$. In the last row we use the solution of equation 
$$
\delta {\cal G} - J=0
$$ 
in the form
$$
{\cal G} = \Sigma[J] + * d \Phi
$$
where $\Sigma$ is the surface spanned on contour $J$.
One can see that integral over $\Phi$ in Eq. (\ref{Phi}) decouples, and we arrive at
\begin{eqnarray}
	Z[{J}]  &=& const \, e^{-i \langle  \Sigma[J],* J \rangle \frac{1}{2\theta}  }\label{Phi2}
\end{eqnarray}
The term in exponent $\langle  \Sigma[J],* J \rangle $ is the integer linking number of the electron worldline with itself. Therefore, we arrive at conclusion that $Z[J]$ does not depend on $J$ for $\theta = 1/{2\pi 2s}$, where $s$ is integer (we assume periodic boundary conditions, so that the boundary terms are absent). This proves the equivalence of the initial model with partition function of Eq. (\ref{Z0}) and the model by Lopez and Fradkin.

\section{The effect of fluctuations of statistical gauge field above its mean value }

\label{AppCSfluct}

In the present section we also will use the calculus of differential forms in flat space - time  (with the real - valued coefficients). The corresponding notations are summarized in Appendix \ref{SectDiff}.

Here we consider the effect of the fluctuations of statistical gauge field around its mean value on the electron density.  We use the "polymer" representation of the theory of Lopez and Fradkin given in Appendix \ref{AppDual}. We separate the statistical field into two contributions: the mean field value ${\cal A}_0$ (its effect on the electrons is taken into account non - perturbatively), and the fluctuations ${\cal A}'$ to be taken perturbatively. We consider the theory in the reference frame without electric field. The Chern - Simons action receives the form (${\cal A}_{0,0} = 0$) 
\begin{equation}
	S_{CS}[{\cal A}_0+{\cal A}']  = S_{CS}[{\cal A}'] + {\theta} \int d^3 z {\cal A}'_t {B}\frac{2sp}{1+2sp}=\frac{\theta}{2}\langle A', *dA'\rangle + \langle A', *\bar{J}_0 \rangle,
\end{equation}
where $\bar{J}_0 = (1,0,0)\, \frac{p}{1+2sp}\, \frac{B}{2\pi}$ is the mean field $3$ - current in the given reference frame. 
 The considered theory has the partition function
 \begin{equation}
 	Z = \int D\psi D\bar{\psi} D \lambda D{\cal A}'  e^{i \frac{\theta}{2}\langle {\cal A}', *d{\cal A}'\rangle + i\langle {\cal A}', *\bar{J}_0 \rangle  + iS_g[\lambda]  + i\int d^3 x \bar{\psi} (\hat{Q}[{\cal A}_0+{\cal A}',\lambda]) \psi} \delta (\delta {\cal A}')
 \end{equation}
 with 
 \begin{eqnarray}
 	S_g & = &  - \frac{1}{4 e^2}\mz{\int d^4z} (\partial_{[\mu} (\lambda)_{\nu]})^2 + \mz{\int d^4z}L_{gauge\, fixing}[\lambda]
\end{eqnarray}
and 
\begin{eqnarray}
\hat{Q}[{\cal A}_0+{\cal A}',\lambda] &=& i \gamma^0 \partial_t -\gamma^0 {\cal A}'_t + \mu\gamma^{\nu} u_{\nu} - \gamma^{\mu} \lambda_\mu(z)+ \gamma^k  \Bigl( i {\partial}_k + B_{eff}x\hat{y}_k - {\cal A}'_k\Bigr) - M  
\end{eqnarray}
where $u = (1,0,-E/B)$ in the {\it given} reference frame {(the moving frame with zero $E$)}. We added the gauge fixing condition for ${\cal A}'$,  which assumes the Lorentz gauge $\delta {\cal A}'=0$. In this gauge we may invert operator $*d$, which gives $*\square^{-1}d$. Integrating out ${\cal A}'$ we arrive at 
\begin{equation}
	Z = const \int D\psi D\bar{\psi} D \lambda   e^{ iS_g[\lambda]  + i\int d^3 x \bar{\psi} (\hat{Q}[{\cal A}_0,\lambda]) \psi - \frac{i}{2\theta} \langle \lrcorner(\bar{\psi}\gamma \psi - \bar{J}_0), *\square^{-1} d (\bar{\psi}\gamma \psi - \bar{J}_0)\rangle} 
\end{equation}
Since $d {\bar J}_0 =0$ we can omit $\bar{J}_0$ in the above expression. Restoring the integration over ${\cal A}'$ we come to 
 \begin{equation}
 	Z = \int D\psi D\bar{\psi} D \lambda D{\cal A}'  e^{i \frac{\theta}{2}\langle {\cal A}', *d{\cal A}'\rangle   + iS_g[\lambda]  + i\int d^3 x \bar{\psi} (\hat{Q}[{\cal A}_0+{\cal A}',\lambda]) \psi} \delta (\delta {\cal A}')
 \end{equation}
Next, as in Appendix \ref{AppDual} we come to the "polymer representation":  
\begin{eqnarray}
	Z &=& \int D \lambda \sum_{n}\Pi_{i = 1... n} (-1)^n Dz^{(i)}  e^{
		iS_g[\lambda] + i\sum_{j = 1...n}\int_{s = 0}^{1} ( A_\mu dz^{(j)\mu}(s) + {\cal A}_{0,\mu} dz^{(j)\mu}(s)+ \lambda dz^{(j)0}+{\cal L}[z^{(j)}(s)]ds) + iS_{int}[z(s)]}Z_{{\cal A}'}[J],\nonumber\\ Z_{{\cal A}'}[J]&=&\int D{\cal A}' \,e^{i S_{CS}[{\cal A}']  + i\sum_{j = 1...n}\int_{s = 0}^{1} {\cal A}'_\mu dz^{(j)\mu}(s)}\delta (\delta {\cal A}')
	\label{Z1pf}
\end{eqnarray}
We omit here the gauge fixing term for ${\cal A}'$, follow the same steps as in Appendix \ref{AppDual}, and have 
\begin{eqnarray}
	Z_{{\cal A}'}[{J}] &=& \int D{\cal A}' \,e^{i \frac{\theta}{2} \langle {\cal A}', *{\cal F}'\rangle +  i\langle {\cal A}', J \rangle}\nonumber\\ &=& \int D{\cal A}' D\Lambda  \,e^{i \frac{\theta}{2} \langle \square^{-1}\delta{\cal F}+d\alpha, *{\cal F}\rangle + i\langle {\cal A}', J\rangle}\delta(\Lambda - {\cal F})
	\nonumber\\ &=& \int D{\cal A}' D\Lambda \,e^{i \frac{\theta}{2} \langle {\cal F}, \square^{-1} d*{\cal F}\rangle + i\langle {\cal A}', J\rangle}\delta(\Lambda - d{\cal A}')
	\nonumber\\ &=& \int D{\cal A}' D\Lambda D {\cal G}\,e^{i \frac{\theta}{2} \langle \Lambda, \square^{-1} d*{\Lambda}\rangle + i\langle {\cal A}', J\rangle + i \langle (\Lambda - d{\cal A}'), {\cal G}\rangle}
	\nonumber\\ &=& \int D{\cal A}' D\Lambda D {\cal G}\,e^{i \frac{\theta}{2} \langle (\Lambda+d*{\cal G}/\theta), \square^{-1} d*{(\Lambda+d*{\cal G}/\theta)}\rangle - \frac{i}{2\theta}\langle {\cal G}, *\delta  {\cal G}\rangle+ i\langle {\cal A}', J\rangle - i \langle   d{\cal A}', {\cal G}\rangle}
	\nonumber\\ &=& const \int   D {\cal G}\,e^{-i \langle {\cal G},*\delta  {\cal G}\rangle \frac{1}{2\theta}  }\delta(\delta {\cal G} - J)\nonumber\\ &=& {const}\,\int   D \Phi\,e^{-i \langle *d\Phi + \Sigma[J], *\delta(* d \Phi + \Sigma[J]) \rangle \frac{1}{2\theta}  }\label{Phif}
\end{eqnarray}
In the last row we use the general solution of equation 
$$
\delta {\cal G} - J=0
$$ 
in the form
$$
{\cal G} = \Sigma[J] +  * d \Phi
$$
where $\Sigma$ is the surface spanned on contour $J$.
As above, the integral over $\Phi$ in Eq. (\ref{Phi}) decouples, and we arrive at
\begin{eqnarray}
	Z_{{\cal A}'}[{J}]  &=& const \, e^{-i \langle  \Sigma[J],* J \rangle \frac{1}{2\theta}  } = const  \label{Phi2f}
\end{eqnarray}
One can see that $Z_{{\cal A}'}[{J}] $ does not depend on $J$, which means that on the level of perturbation theory there are no corrections to the partition function due to the fluctuations of the statistical gauge field around its mean value.

\section{Derivation of the expression for the Hall conductivity through matrix
elements of the effective mean field Hamiltonian}

\label{app0}

In the theory by Lopez and Fradkin, in the mean field approximation $\mathbf{G}$ is
equal to the free fermion Green function in the presence of the effective
magnetic field and external electric field. The external electric field is defined through an addition $%
A_0(x) = - E x$ to the Hamiltonian, i.e. as an additon to $Q$. The electric
current in this approximation is given by 
\begin{equation}
j^k = \mathrm{Tr}\, \int \frac{d^3 p}{(2\pi)^3} {({G}_{eff})}_W(x,p)
\partial_{p^k} (Q_{eff})_W(x,p)
\end{equation}
The response of the electric current to the external electric field is reduced to
the response of ${G}_{eff}$ to it. $G_{eff, W}$ is defined as solution of the
Groenewold equation 
\begin{align}
G_{eff, W} \star Q_{eff, W} = 1
\end{align}
Its solution may be represented as a series in gradient expansion. The term
linear in electric field results in 
\begin{equation}
{G}^{(1)}_{ W} = -\frac{i}{2} (G_{eff})_W(p,x)\star \frac{\partial
(Q_{eff})_W(p,x)}{\partial p_i} \star (G_{eff})_W(p,x)\star \frac{\partial
(Q_{eff})_W(p,x)}{\partial p_j} \star (G_{eff})_W(p,x) \, F_{ij}
\end{equation}
Here on the right hand side we use $Q_{eff}$ and $G_{eff}$ with vanishing electric field.
We choose the gauge in which the gauge potential is 
\[
A_x = 0, \quad A_y = B x\,. 
\]
The external electric field $\mathit{E}_y$ corresponding to the gauge potential $%
A$ is directed along the $y$ axis. The above derived expressions give the
following result for the electric current averaged over the area of the
system: 
\begin{equation}
\langle j_{x} \rangle \approx  \frac{\mathcal{S}}{2\pi}\mathcal{N} \mathit{E_{eff, y}}\,.
\end{equation}
Here $E_{eff} = E/(1+2sp)$.
The average Hall conductivity is $\sigma_{xy} = \frac{\mathcal{N}}{2 \pi} \times \frac{1}{2sp+1}$ (recall
that $j_x = +\sigma_{xy} E_y$) while $\mathcal{N}$ is the following
expression ($A=0$), 
\begin{eqnarray}
\mathcal{N} & = & {\frac{T}{3! 4\pi^2 \mathcal{S}}} \, \int \,\mathrm{Tr}\,
d^3p \, d^3x\, \epsilon_{ijk}\, \Big[ (G_{eff})_W(p,x)\star \frac{\partial
(Q_{eff})_W(p,x)}{\partial p_i}  \nonumber \\
&& \star (G_{eff})_W(p,x)\star \frac{\partial (Q_{eff})_W(p,x)}{\partial p_j}
\star (G_{eff})_W(p,x)\star \frac{\partial (Q_{eff})_W(p,x)}{\partial p_k} %
\Big]_{E=0}\,.  \label{calM2d23_}
\end{eqnarray}
where $(Q_{eff})_W (p,x)$ is the Wigner transform of $Q_{eff} (p,x)$, with 
\[
Q_{eff}(p,x) = i\omega - H_{eff}(p_x,p_y - B_{eff} x).
\]
and $(G_{eff})_W$ is the Green's function calculated as a inverse of $(Q_{eff})_W$. Here, we include the effect of interactions through the
effective magnetic field $B_{eff} = B - \rho_0 / \theta$ \cite{Fradkin1991chern}.

%Eq. (\ref{calM2d23_}) can be derived as follows. 

\section{Expression for $\mathcal{N}$ through the effective Hamiltonian}

\label{app1}

We proceed now to calculate $\mathcal{N}$ explicitly within the mean field approximation. If we expand $\epsilon_{\mu\nu\rho}$ for the temporal component we have 
\begin{align}  \label{N-2}
\mathcal{N}=&{\frac{T}{8\pi^2 \mathcal{S}}} \, \int\, d^3p \, d^3x\, \epsilon_{ij}\cdot\nonumber\\
&Tr\Big[ (G_{eff})_W \star \partial_{p_3} (Q_{eff})_W \star (G_{eff})_W\star
\partial_{p_i} (Q_{eff})_W \star (G_{eff})_W \star \partial_{p_j} (Q_{eff})_W \nonumber \\
&-(G_{eff})_W \star \partial_{p_i} (Q_{eff})_W \star (G_{eff})_W\star\partial_{p_3} (Q_{eff})_W \star (G_{eff})_W \star \partial_{p_j} (Q_{eff})_W
\Big].  \nonumber \\
\end{align}
In Wigner representation, the quantity $\partial_{p_i} (Q_{eff})_W$ is calculated as follows
\begin{eqnarray}
\frac{\partial}{\partial p^i}(Q_{eff})_W(x,p) &=& \frac{\partial}{\partial
p^i}\int \frac{d^3P}{(2\pi )^3}~e^{iPx} \tilde{Q}_{eff}\Big(p+\frac{P}{2},p-\frac{P}{2}\Big)  \nonumber
\\
&=& \int \frac{d^3P}{(2\pi )^3}~e^{iPx}\Big(\frac{\partial}{\partial K_1^i }+\frac{\partial}{%
\partial {K_2^i} }) \tilde{Q}_{eff}(K_1,K_2)\Big|_{K_1=p+\frac{P}{2}}^{K_2=p-\frac{P%
}{2}}\, 
\end{eqnarray}
where we used the chain rule of differentiation to get 
\begin{align}
\frac{\partial}{\partial {p^i} } =\frac{\partial K_1^j}{\partial {p^i} }%
\frac{\partial}{\partial {K_1^j} }+\frac{\partial K_2^j}{\partial {p^i} }%
\frac{\partial}{\partial {K_2^j} } =\frac{\partial}{\partial {K_1^i} }+\frac{%
\partial}{\partial {K_2^i} }.
\end{align}
If we denote 
\begin{align}
\tilde{Q}_{eff,i}\Big(p+\frac{P}{2},p-\frac{P}{2}\Big) = \Big( \frac{\partial}{\partial {K_1^i}} +\frac{\partial}{\partial{K_2^i}}\Big) 
\tilde{Q}_{eff}(K_1,K_2)\Big|_{K_1=p+P/2}^{K_2=p- P/2}
\end{align}
then $\partial_{p_i}(Q_{eff})_W$ can be written in the short-hand form as $\partial_{p_i}(Q_{eff})_W=(Q_{eff,i})_W$. Using the associativity of the Moyal star product the first term of Eq. (\ref{N-2}) is written as 
\begin{eqnarray}
&&(G_{eff})_W \star \partial_{p_3} (Q_{eff})_W \star (G_{eff})_W\star
\partial_{p_j} (Q_{eff})_W \star (G_{eff})_W \star \partial_{p_k} (Q_{eff})_W 
\nonumber \\
&&=\int \frac{d^3P}{(2\pi )^3}~e^{iPx}(\tilde{G}_{eff}\tilde{Q}_{eff,3}\tilde{G}_{eff}\tilde{Q}_{eff,j}\tilde{G}_{eff}\tilde{Q}_{eff,k})
\Big(p+\frac{P}{2},p-\frac{P}{2}\Big)
\end{eqnarray}
where 
\begin{eqnarray}
&& (\tilde{G}_{eff}\tilde{Q}_{eff,3}\tilde{G}_{eff}\tilde{Q}_{eff,j}\tilde{G}_{eff}\tilde{Q}_{eff,k})\Big(p+\frac{P}{2},p-\frac{P}{2}\Big)  \nonumber \\
&& = \int d^3p^{(2)}d^3p^{(3)}d^3p^{(4)}d^3p^{(5)}d^3p^{(6)}\cdot  \nonumber \\
&& \ \ \ \ \Big[ \tilde{G}_{eff}(p + \frac{P}{2},p^{(2)})\Big( \lbrack\partial_{p^{(2)}_3} + \partial_{p^{(3)}_3}] \tilde{Q}_{eff}(p^{(2)},p^{(3)})\Big)\cdot \nonumber \\
&& \ \ \ \ \tilde{G}_{eff}(p^{(3)},p^{(4)})\Big( \lbrack \partial_{p^{(4)}_j} +\partial_{p^{(5)}_j}] \tilde{Q}_{eff}(p^{(4)},p^{(5)})\Big)\cdot  \nonumber \\
&& \ \ \ \ \tilde{G}_{eff}(p^{(5)},p^{(6)})\Big( \lbrack \partial_{p^{(6)}_k} +\partial_{p^{(7)}_k}] \tilde{Q}_{eff}(p^{(6)},p^{(7)})\Big) \Big|_{p^{(7)}=p-\frac{P}{2}} \Big].  
\label{GQ0GQiGQj}
\end{eqnarray}
The second term of Eq. (\ref{N-2}) can also be expressed in a similar way. Substituting the above expressions back into Eq. (\ref{N-2}) and integrating over $x$ and $P$ leads to ($T$ vanishes due to imaginary time independence of the propagator (and its inverse) and integration over $d\tau$)
\begin{align}
\mathcal{N}=&{\frac{1}{8\pi^2\mathcal{S}}} \int d^3p^{(1)}d^3p^{(2)}d^3p^{(3)}d^3p^{(4)}d^3p^{(5)}
d^3p^{(6)} \epsilon_{ij}\, \cdot  \nonumber \\
&Tr\Big[\tilde{G}_{eff}(p^{(1)} ,p^{(2)})\Big( \lbrack\partial_{p^{(2)}_3} + \partial_{p^{(3)}_3}] \tilde{Q}_{eff}(p^{(2)},p^{(3)})\Big)\cdot \nonumber \\
&\tilde{G}_{eff}(p^{(3)},p^{(4)})\Big( \lbrack \partial_{p^{(4)}_i} +\partial_{p^{(5)}_i}] \tilde{Q}_{eff}(p^{(4)},p^{(5)})\Big)\cdot  \nonumber \\
&\tilde{G}_{eff}(p^{(5)},p^{(6)})\Big( \lbrack \partial_{p^{(6)}_j} +\partial_{p^{(1)}_j}] \tilde{Q}_{eff}(p^{(6)},p^{(1)})\Big)   \nonumber \\
&-\tilde{G}_{eff}(p^{(1)} ,p^{(2)})\Big( \lbrack \partial_{p^{(2)}_i} +\partial_{p^{(3)}_i}] \tilde{Q}_{eff}(p^{(2)},p^{(3)})\Big)\cdot  \nonumber \\
&\tilde{G}_{eff}(p^{(3)},p^{(4)})\Big( \lbrack \partial_{p^{(4)}_3} +\partial_{p^{(5)}_3}] \tilde{Q}_{eff}(p^{(4)},p^{(5)})\Big)\cdot  \nonumber \\
&\tilde{G}_{eff}(p^{(5)},p^{(6)})\Big( \lbrack \partial_{p^{(6)}_j} +\partial_{p^{(1)}_j}] \tilde{Q}_{eff}(p^{(6)},p^{(1)})\Big)\Big].
\label{GQ0GQiGQj-2}
\end{align}
The momentum space function $\tilde{Q}_{eff}(p^{(i)},p^{(j)})$ has the representation 
\begin{equation}
\tilde{Q}_{eff}(p^{(i)},p^{(j)}) = \langle p^{(i)}| \hat{Q}_{eff} | p^{(j)}\rangle =
\Big(  i\omega^{(i)}\delta^{2} (p^{(i)} - p^{(j)}) - \langle p^{(i)} |
\hat{H}_{eff} | p^{(j)} \rangle \Big) \delta (\omega^{(i)}-\omega^{(j)})  \label{Q1}
\end{equation}
where $p = (p_3,p_1,p_2) = (\omega, \mathbf{p})$. The corresponding Green function can be calculated as 
\begin{align}
\tilde{G}_{eff}(p^{(i)},p^{(j)}) = \sum_{n} \frac{1}{i\omega^{(i)} - \mathcal{E}_n}
\langle {p}^{(i)}| n \rangle \langle n | p^{(j)}\rangle
\delta(\omega^{(i)}-\omega^{(j)})\,. 
\end{align}
The numbers $\mathcal{E}_n$ represent the energy eigenvalues of $\hat{H}_{eff}$ (shifted by the chemical potential). From Eq. (\ref{Q1}) it may be concluded that 
\begin{eqnarray}
\partial_{p^{(i)}_3} \tilde{Q}_{eff} (p^{(i)},p^{(i+1)}) &=& i\delta^{2} (p^{(i)}
- p^{(i+1)})\delta (\omega^{(i)}-\omega^{(i+1)})  \label{p0' Q} \\
(\partial_{p^{(i)}_j} + \partial_{p^{(i+1)}_j}) \tilde{Q}_{eff} (p^{(i)},p^{(i+1)})
&=& -(\partial_{p^{(i)}_j} + \partial_{p^{(i+1)}_j}) \langle p^{(i)} |
\hat{H}_{eff} | p^{(j)} \rangle \delta(\omega^{(i)}-\omega^{(i+1)})  \nonumber \\
& = & -i \langle p^{(i)} | \hat{H}_{eff} \hat x_j - \hat x_j \hat{H}_{eff} | p^{(i+1)}
\rangle \delta(\omega^{(i)}-\omega^{(i+1)})  \nonumber \\
& = & -i \langle p^{(i)} | [\hat{H}_{eff}, \hat{x}_j]| p^{(i+1)}\rangle
\delta(\omega^{(i)}-\omega^{(i+1)}).
\label{pro-Q}
\end{eqnarray}
We can easily integrate out all intermediate $\omega^{(i)}$'s by means of the delta functions to remain with a single $\omega$ and also integrating
out $(p^{(2)},p^{(3)})$ of the first term and $(p^{(4)},p^{(5)})$ of the second term of Eq. (\ref{GQ0GQiGQj-2}) after substituting the expression of Eq. (\ref
{p0' Q}) and using the completeness relation $\hat{1}=\int d^3p_i|p_i \rangle \langle p_i |$. After such operations Eq. (\ref{GQ0GQiGQj-2}) reduces to 
\begin{align}
\mathcal{N}=&\frac{i}{8\pi^2\mathcal{S}}\,\sum_{n,k} \int d\omega d^2p^{(1)}d^2p^{(2)}d^2p^{(3)}d^2p^{(4)} \epsilon_{ij}\cdot  \nonumber\\
&\mathrm{Tr}\,\Big[ \frac{1}{(i\omega^{} - \mathcal{E}_n)^2} \langle {\ p}^{(1)}| n \rangle \langle n | {\ p}^{(2)}\rangle \langle {p}^{(2)}| [\hat{H}_{eff}, \hat x_i] | {\ p}^{(3)}\rangle \frac{1}{(i\omega^{} - \mathcal{E}_k)}\langle {p}^{(3)}| k \rangle \langle k | {\ p}^{(4)}\rangle \langle {\ p}
^{(4)}| [\hat{H}_{eff}, \hat x_j] | {p}^{(1)}\rangle  \nonumber \\
& -\frac{1}{(i\omega^{} - \mathcal{E}_k)^2} \langle {\ p}^{(1)}| k \rangle\langle k | {\ p}^{(2)}\rangle \langle {p}^{(2)}| [\hat{H}_{eff}, \hat x_i] | {\
p}^{(3)}\rangle \frac{1}{(i\omega^{} - \mathcal{E}_n)} \langle {p}^{(3)}| n\rangle \langle n | {\ p}^{(4)}\rangle \langle {\ p}^{(4)}| [\hat{H}_{eff}, \hat
x_j] | {p}^{(1)}\rangle \Big]. 
\end{align}
where we redefine the momentum variables left, as $p^{(1)}, p^{(2)},p^{(3)}, p^{(4)}$. Further integrating out the momenta $p^{(i)}$ we have 
\begin{eqnarray}  
\label{N-3}
\mathcal{N} &=& \frac{i}{8\pi^2\mathcal{S}}\,\sum_{n,k} \int \,d \omega \,
\epsilon_{ij}\,\Big[ \frac{1}{(i\omega^{} - \mathcal{E}_n)^2} \langle n| [\hat{
H}_{eff}, {\hat x}_i] | k \rangle \frac{1}{(i\omega^{} - \mathcal{E}_k)}
\langle k | [\hat{H}_{eff}, {\hat x}_j] | n \rangle  \nonumber \\
&& - \frac{1}{(i\omega^{} - \mathcal{E}_k)^2} \langle k| [\hat{H}_{eff}, {\hat
x}_i] | n \rangle \frac{1}{(i\omega^{} - \mathcal{E}_n)} \langle n | [\hat{
H}_{eff}, {\hat x}_j] | k \rangle \Big]  \nonumber \\
&=& - \frac{i}{2\pi \mathcal{S}}\,\sum_{n,k} \epsilon_{ij}\, \langle n | [\hat{
H}_{eff}, \hat x_i] | k \rangle \langle k | [\hat{H}_{eff}, \hat x_j] | n
\rangle \frac{ (\Theta(- \mathcal{E}_n) \Theta(\mathcal{E}_k) - \Theta(- 
\mathcal{E}_k) \Theta(\mathcal{E}_n))}{(\mathcal{E}_k - \mathcal{E}_n)^2}. 
\end{eqnarray}
In the last expression we have exchanged the notations $n \rightarrow k, k \rightarrow n$ for the second term in order to write it in a compact form.\par
Now, we will define new variables $(\xi_i, X_i)$ as follows 
\begin{align}
\hat{x}_1 = -\frac{\hat{p}_y - B_{eff} x}{B_{eff}} + \hat{X}_1 = \hat{\xi}_1+ \hat{X}_1,\,\,\,\,\hat{x}_2 = \frac{\hat{p}_x}{B_{eff}} + \hat{X}_2= \hat{\xi}_2 + \hat{X}_2\,. 
\end{align}
The commutation relations implied by the above set of variables are 
\begin{align}
[\hat{\xi}_{i},\hat{\xi}_{j}] = \frac{i}{B_{eff}}\epsilon_{ij}, \,\,\,\, [\hat{X}_{i},\hat{X}_{j}] = - \frac{i}{ B_{eff}}\epsilon_{ij},\,\,\,\,
[\hat{H}_{eff}, \hat{X}_{i}] = 0 \,\,\,\,\Big(\hat{H}_{eff}=\frac{B_{eff}^2}{2m}(\hat{\xi}_1^2+\hat{\xi}_2^2)\Big)
\end{align}
where $i$ labels the first and second set of variables, respectively. In terms of the $\hat{\xi}_i$'s, Eq. (\ref{N-3}) can be written as 
\begin{eqnarray}
\mathcal{S}\mathcal{N} = -\frac{i}{2\pi} \sum_{n,k} \epsilon_{ij} \,\Big[ \langle n| [\hat{H}
_{eff}, {\hat \xi}_i] | k \rangle \langle k | [\hat{H}_{eff}, {\hat \xi}_j] | n
\rangle \Big] \frac{( \Theta(-\mathcal{E}_n)\Theta(\mathcal{E}_k) - \Theta(-%
\mathcal{E}_k)\Theta(\mathcal{E}_n))}{(\mathcal{E}_k - \mathcal{E}_n)^2}.
\end{eqnarray}

\section{Calculation of $\mathcal{N}$}
\label{app2}

The two terms involving Heaviside step functions entering with opposite signs give identical contributions to $A\mathcal{N}$ such that
\begin{align}
\frac{1}{2}\mathcal{S}\mathcal{N}&=-\frac{i}{2\pi}
\sum_{n,k}\epsilon_{ij}\langle n| [\hat{H}_{eff},\hat{\xi}_i]| k\rangle\langle
k|[\hat{H}_{eff},\hat{\xi}_j]|n\rangle \frac{\Theta (-\mathcal{E}_n)\Theta (\mathcal{E}_k)}{(\mathcal{E}_n-\mathcal{E}_k)^2}  \nonumber \\
&=-\frac{i}{2\pi} \sum_{n,k}\epsilon_{ij}\langle n|\hat{\xi}_i| k\rangle\langle k|\hat{\xi}_j|n\rangle \Theta (-\mathcal{E}_n)\Theta (\mathcal{E}_k)  \nonumber\\
&=-\frac{i}{2\pi} \sum_{n,k}\epsilon_{ij}\langle n|\hat{\xi}_i| k\rangle\langle k|\hat{\xi}_j|n\rangle \Theta (-\mathcal{E}_n)(\Theta (\mathcal{E}_k)+\Theta (-\mathcal{E}_k))  \nonumber \\
&=-\frac{i}{2\pi} \sum_{n,k}\epsilon_{ij}\langle n|\hat{\xi}_i| k\rangle\langle k|\hat{\xi}_j|n\rangle \Theta (-\mathcal{E}_n)  \nonumber \\
&=-\frac{i}{2\pi} \sum_{n}\langle n|[\hat{\xi}_1,\hat{\xi}_2]|n\rangle \Theta (-\mathcal{E}_n)  \nonumber \\
&=-\frac{i}{2\pi} \sum_{n}\langle n|\frac{i}{B_{eff}}|n\rangle \Theta (-\mathcal{E}_n)\nonumber \\
&=\frac{1}{2\pi}\int dp_{y}\frac{1}{B_{eff}}L\sum_{m}\Theta (-\mathcal{E}_m)  \nonumber \\
&=\frac{1}{2}\mathcal{S}p
\end{align}
for $p$ filled effective Landau levels resulting in $\mathcal{N}=p$ and thereby an effective IQHE as expected. Notice that the index labeling energy eigenvalues $n=(p_y,m)$ splits into the discrete index $m$ labeling Landau levels and the continuous index $p_y$ labeling intra Landau level states. Furthermore our momentum space scaling and the Landau level degeneracy allow us to reach the ultimate equality from the penultimate one, since the momentum range is given by 
\begin{align}
\Delta p_y=2\pi \frac{L}{l_{B_{eff}}^2},\,\,\,\, l_{B_{eff}}^2=(B_{eff})^{-1}.
\end{align}
$l_{B_{eff}}$ is the effective magnetic length. It follows from the above calculation that $\mathcal{N}$ counts the number of fully occupied Landau levels in the mean field composite fermion theory within the FQHE. Analogously it counts the number of fully occupied Landau levels in the theory of free electrons within the IQHE.

%%%%%%%%%%%%%%%%%%%%%%%%%%%%%%%%%%%%%%%%%%%%%%%%%%%%%%

%%%%%%%%%%%%%%%%%%%%%%%%%%%%%%%%%%%%%%%%%%%%%%%%%%%%%%%%%%%%%%%%%%%%%%%%%%%%%%%%%%%%%%%%%%%%%%%%%%%%%%%%%%%%
%%%%%%%%%%%%%%%%%%%%%%%%%%%%%%%%%%%%%%%%%%%%%%%%%%%%%%%%%%%%%%%%%%%%%%%%

\section{Feynman rules for the Wigner - transformed Green functions}
\label{sectbubble}

In this section we briefly remind the reader the Feynman rules written in terms of Wigner - transformed Green functions (for more details see \cite{Zhang2022}). Let us consider the Dirac operator of the form of Eq. (\ref{QE0}):
\begin{eqnarray}
	\hat{Q}[{\cal A},\lambda] &=& i \gamma^0 \partial_t -\gamma^0 {\cal A}'_t + \mu\gamma^{\nu} u_{\nu} - \gamma^{\mu} \lambda_\mu(z)+ \gamma^k  \Bigl( i {\partial}_k + B_{eff}x\hat{y}_k - {\cal A}'_k\Bigr) - M  ,
\end{eqnarray}
where $u = (1,0,-E/B)$. The two - point Green function matrix element $G$ satisfies the equation $\int dxQ(x_1,x) G(x,x_2)=\delta(x_1-x_2)$. With the definition of Wigner - Weyl transform, $G$ is written as
\begin {eqnarray}\label{Wigner}
G_W (x,p)=\int dr G(x+r/2, x-r/2) e^{-ipr}, \nonumber \\
\end{eqnarray}
for which the Groenewold equation reads $Q_W(x,p)\star G_W (x,p)=1,$ where $Q_W$ is Weyl symbol of the above Dirac operator $\hat{Q}$, while the Moyal product is defined as $\star=e^{i(\overleftarrow\partial_x\overrightarrow\partial_p - \overleftarrow\partial_p\overrightarrow\partial_x) /2}$.

In Fig. (\ref{fig.1}) the lowest order Feynman diagram for the self-energy is shown. Fermionic propagators are represented by solid lines, while the photon propagator $D_W^{\mu_1 \mu_2}(x,p)$ is represented by a curvy line. It is also supposed here that the Wigner - transformed photon propagator depends both on $x$ and on $p$, although in practical applications to the main text we consider the simpler case of $D_W$ depending only on $p$.

\begin{figure}[h]
	\centering  
	\begin{equation}
		\begin{tikzpicture}
			\begin{feynman}
				\vertex (a1);
				\vertex [right=1.0cm of a1] (a2);
				\vertex [right=3.0cm of a2] (a3);
				\vertex [right=1.0cm of a3] (a4);
				\feynmandiagram [horizontal= a1 to a4 ] {
					(a1) -- [edge label'=\( G_W \)] (a2) -- [edge label'=\( G_W \)] (a3) -- [edge label'=\( G_W \)] (a4),
					(a2) -- [photon, half left, edge label'=\( D_W^{\mu_1 \mu_2} \)]  (a3), }; 
			\end{feynman}
		\end{tikzpicture}	
	\end{equation}
	\caption{The schematic representation of the diagrams of the fermionic self - energy without internal fermion loops. The solid lines represent the fermionic propagators $G_W(x,p) $ and the curvy line represents the photon propagator $D_W^{\mu_1 \mu_2}$.}  
	\label{fig.1}   
\end{figure}
The diagram rules can be better understood if we consider the particular examples of Fig.\ref{fig.2}. Indeed, Fig.\ref{fig.2}(a) represents a one - loop interaction diagram. Its expression is given by

\begin {eqnarray}\label{Green_1st}
\int \Big[ G_W(x,p) \gamma^{\mu_1}  \circ_1 \star G_W(x,p-k)\gamma^{\mu_2}\circ_1 \star G_W(x,p)\Big]  %\nonumber \\
D^{(1)\mu_1 \mu_2}_W(x,k) dk. \nonumber
\end{eqnarray}
%%%%%%%%%%%%%%%%%%%%%%%%%%%
We define $\circ_j=e^{-i\overleftarrow\partial_p\partial^{(j)}_x /2}$ and
${}_j\circ =e^{i \partial^{(j)}_x \overrightarrow\partial_p /2}$ as left- and right - sided operators acting on the fermion propagators, respectively, while $\partial_x^{(j)}$ acts on $D^{(j)}$ only. Indeed, the operator  $\overrightarrow\partial_p$ acts on all fermion Green functions standing right to the symbol ${}_j\circ$. The operator   $\overleftarrow\partial_p$ acts on all fermion Green functions standing left to the symbol $\circ_j$.  With these notations, the second diagram of Fig.\ref{fig.2} can be described as  
\begin {eqnarray}\label{GreenStar_eg}
\int\int && [G_W(x,p) \gamma^{\mu_1}\circ_1 \star G_W(x,p-k_1)\gamma^{\mu_3}\circ_2 \star %\nonumber \\
G_W(x,p-k_1-k_2)\gamma^{\mu_2}\star {}_{1}\circ    \nonumber\\
&&G_W(x,p-k_2)\star \gamma^{\mu_4}  {}_{2}\circ G_W(x,p)]            % \nonumber\\
D^{(1)\mu_1 \mu_2}_W(x,k_1) D^{(2)\mu_3 \mu_4}_W(x,k_2) dk_1 dk_2.  \nonumber
\end{eqnarray}

\begin{figure}[h]
\centering  %
\begin{equation}
	\begin{tikzpicture}
		\begin{feynman}
			\vertex (a1);
			\vertex [right=1.0cm of a1] (a2);
			\vertex [right=2.0cm of a2] (a3);
			\vertex [right=1.0cm of a3] (a4);
			\vertex [right=0.5cm of a4] (a5);
			\vertex [right=1.0cm of a5] (a6);
			\vertex [right=1.0cm of a6] (a7);
			\vertex [right=1.0cm of a7] (a8);
			\vertex [right=1.0cm of a8] (a9);
			\vertex [right=1.0cm of a9] (a10);
			\diagram* [horizontal= a1 to a10 ] {
				(a1) -- (a2) -- [edge label'=\( (a) \)] (a3) -- (a4),
				(a5) --  (a6), (a6) -- [edge label'=\( (b)\)] (a7), (a7) -- (a10), 
				(a2) -- [photon, half left, edge label'=\( D^{\mu_1 \mu_2} \)] (a3), 
				(a6) -- [photon, half left, edge label'=\( D^{\mu_1 \mu_2} \)] (a8), 
				(a7) -- [photon, counterclockwise, half right, edge label'=\( D^{\mu_3 \mu_4} \)] (a9),
			}; 
		\end{feynman}
	\end{tikzpicture}	\nonumber
\end{equation}
\caption{(a) One - loop Feynman diagram for the self - energy, 
(b) An entangled two - loop Feynman diagram for the self - energy.}  %
\label{fig.2}   %
\end{figure}
In the cases where the fermion line closes, one has to consider the bubble diagrams represented in Fig. \ref{fig.3}. The first diagram reads
\begin{eqnarray}\label{bubble_a}
\frac{1}{2}\int Tr \Big[ G_W(x,p-k) \gamma^{\mu_1 } \star {}_{1}\circ G_W(x,p)\gamma^{ \mu_2}\Big]%\nonumber \\
  D^{(1)\mu_1 \mu_2}_W(x,k) dk, \nonumber
\end{eqnarray}
which due to properties of the trace reduces to
\begin {eqnarray}\label{bubble_a'}
\frac{1}{2}\int Tr [G_W(x,p)\gamma^{\mu_1} \circ_{1}\star G_W(x,p-k)\gamma^{ \mu_2}]  D^{(1)\mu_1 \mu_2}_W(x,k) dk. \nonumber
\end{eqnarray}
Similarly, the second diagram of Fig. \ref{fig.3} is expressed as
\begin{eqnarray}\label{bubble_b}
&&\frac{1}{4}\int Tr \Big[\gamma^{\mu_1} G_W(x,p-k_1) \gamma^{\mu_3} \circ_2 \star   %\nonumber\\
G_W(x,p-k_1-k_2)    \gamma^{\mu_2} \star {}_1\circ G_W(x,p-k_2)\gamma^{\mu_4}\star {}_2\circ   \nonumber\\
&& G_W(x,p)\Big]       D^{(1)\mu_1 \mu_2}_{W}(x,k_1) D^{(2)\mu_3 \mu_4}_{W}(x,k_2) dk_1 dk_2.
\end{eqnarray}

\begin{figure}[h]
	\centering  %
	\begin{equation}
		\begin{tikzpicture}
			\begin{feynman}
				\vertex (a1);
				\vertex [right=2.0cm of a1] (a2);
				
				\diagram* {
					(a1) -- [half left] (a2), 
					(a1) --  [half right, counterclockwise, edge label'=\( (a)\)](a2),
					(a1) -- [photon] (a2),
				};
			\end{feynman}
		\end{tikzpicture}	
		%\end{equation}
		%\begin{equation}
		\hspace{1.0cm}
		\begin{tikzpicture}
			\begin{feynman}
				\vertex (a1);
				\vertex [right=2.0cm of a1] (a2);
				
				\vertex [below=3.0em of a1] (b1);
				\vertex [below=3.0em of a2] (b2);
				
				\diagram* {
					(a1) -- [quarter left] (a2), 
					(b1) --  [quarter right, counterclockwise, edge label'=\( (b)\)] (b2),
					(a2) -- [quarter left] (b2), 
					(a1) --  [quarter right, counterclockwise](b1),
					(a1) -- [photon] (b2), (a2) -- [photon] (b1),
				};
			\end{feynman}
		\end{tikzpicture}	
	\end{equation}
	\caption{Fermionic bubble diagrams wihtout current insertions at leading and next to leading order}  %
	\label{fig.3}   %
\end{figure}

Now, we will consider the rules for the calculation of Feynman diagrams considered above with the corrections to the fermion propagator in quenched approximation as studied in \cite{Zhang2022}. The only difference is that, instead of the scalar propagator $D^{(j)}(R,k_j)$, here we will consider the photon propagator $D^{(j)\mu_1 \mu_2}_W (x,p)$. We will first present the Feynman diagram expressions in configuration space, since it turns out that concise expressions are obtained, when the photon propagator remains in its configuration space form.

Let us consider the case with one internal fermion loop, whereas the number of external fermion legs is two or zero. If there are $n$ fermionic loop propagators, then there should be a factor $(-1)^n$ coming from the minus sign in the expression of $G(x,y) = - \langle \psi_x \bar{\psi}_y\rangle$. Such a diagram is presented in Fig. \ref{fig.4}. Its form is given by 
\begin {eqnarray}\label{Green_2line(1)}
{\cal F}(x_1|x_2) &=&  %\nonumber\\
\int G(x_1,y_1)\gamma^{\mu_1}G(y_1,y_2)\gamma^{\mu_3}G(y_2,y_3)\gamma^{\mu_5}G(y_3,x_2)
Tr[\gamma^{\mu_2} G(y_4,y_5)\gamma^{\mu_4} G(y_5,y_6)\gamma^{\mu_6}G(y_6,y_4)]    \nonumber\\
&& \quad      D^{\mu_1 \mu_2}(y_1,y_4)D^{\mu_3 \mu_4}(y_2,y_5)D^{\mu_5 \mu_6}(y_3,y_6) dy_1 ... dy_6.
\end{eqnarray}
With the application of the Moyal product $\circ$ between the fermionic and photonic propagators the above expression becomes
\begin {eqnarray}\label{Green_2line(3)}
{\cal F}_W(x_1|p_1) &=&
G_W(x_1,p_1) \gamma^{\mu_1} \star \overset{(1,1)}{\circ} G_W(x_1,p_1)\gamma^{\mu_3} \star  %\nonumber\\
\overset{(2,1)}{\circ} G_W(x_1,p_1)\gamma^{\mu_5} \star \overset{(3,1)}{\circ} G_W(x_1,p_1)  \nonumber\\
&& \int Tr [  \gamma^{\mu_2} \overset{(1,2)}{\circ} G_W(x_2,p_2)\gamma^{\mu_4}  \star \overset{(2,2)}{\circ} G_W(x_2,p_2)\gamma^{\mu_6} \star  %\nonumber\\
\overset{(3,2)}{\circ} G_W(x_2,p_2)]  \nonumber\\
&& \quad    D^{(1)\mu_1 \mu_2}(x_1,x_2)D^{(2)\mu_3 \mu_4}(x_1,x_2) D^{(3)\mu_5 \mu_6}(x_1,x_2) dx_2 dp_2.
\end{eqnarray}
Here \mz{$\overset{(i,j)}{\circ}=e^{\frac{i}{2}(\partial^{(i)}_{x_j}\overrightarrow\partial_{p_j}-\overleftarrow\partial_{p_j}\partial^{(i)}_{x_j} )}$}. %\mzo{Possibly, there is misprint here. Please check.} 

\begin{figure}[h]
\centering  %
\begin{equation}
	\begin{tikzpicture}
		\begin{feynman}
			\vertex (a1);
			\vertex [right=2.0cm of a1] (a2);
			\vertex [right=2.0cm of a2] (a3);
			
			\vertex [below=5.0em of a1] (b1);
			\vertex [below=5.0em of a2] (b2);
			\vertex [below=5.0em of a3] (b3);
			
			\vertex [below=5.0em of b1] (c1);
			\vertex [below=5.0em of b2] (c2);
			\vertex [below=5.0em of b3] (c3);
			
			\vertex [below=1.5em of c1] (d1);
			\vertex [below=1.5em of c2] (d2);
			\vertex [below=1.5em of c3] (d3);
			
			\diagram* {
				(b1) -- [quarter left] (a2), 
				(a2) --  [quarter left](b3),
				(b3) -- [quarter left, counterclockwise] (c2), 
				(c2) --  [quarter left, counterclockwise](b1),
				(d1) -- (d3), 
				(b1) -- [photon, edge label'=\( D^{\mu_1 \mu_2}\)] (d1), 
				(c2) -- [photon, edge label'=\( D^{\mu_3 \mu_4}\)] (d2), 
				(b3) -- [photon, edge label=\( D^{\mu_5 \mu_6}\)] (d3),
			};
		\end{feynman}
	\end{tikzpicture}	
\end{equation}
\caption{An example of a Feynman diagram contributing to the self - energy, which contains two fermion lines. } \label{fig.4}   
\end{figure}
In the case of diagrams with more than two legs we can consider any diagram, but with an even number of external fermion legs. Take for instance the diagram of Fig.\ref{fig.5}. In configuration space representation, this can be written as
\begin {eqnarray}\label{Green_4legs_a}
{\cal F}(x_1,y_1|x_2,y_2)= %\nonumber \\
\int G(x_1,x_1')\gamma^{\mu_1}G(x'_1, y_1)  D^{(1)\mu_1\mu_2}(x_1',x_2') %\nonumber \\
G(x_2,x_2')\gamma^{\mu_2} G(x_2',y_2) dx_1' dx_2',
\end{eqnarray}
which with the application of the Wigner - transformation becomes
\begin {eqnarray}\label{Green_4legs_b}
{\cal F}_W(x_1,x_2|p_1,p_2)=
[G_W(x_1,p_1)\gamma^{\mu_1}\overset{(1,1)}{\circ} \star G_W(x_1,p_1)] %\nonumber \\
[G_W(x_2,p_2)\gamma^{\mu_2}\overset{(1,2)}{\circ} \star G_W(x_2,p_2)]   D^{(1)\mu_1 \mu_2}(x_1,x_2)
\end{eqnarray}

\begin{figure}[h]
\centering  %
\begin{equation}
	\begin{tikzpicture}
		\begin{feynman}
			\diagram* [layered layout, horizontal= {{a to c},{i1 to i3}}] {
				a [particle=\( x_1 \)]--  b [dot] -- c [particle=\( y_1 \)], 
				i1 [particle=\( x_2 \)] -- i2 [dot] -- i3 [particle=\( y_2 \)], b -- [photon, edge label'=\( D^{\mu\nu } \)] i2 , 
			};
		\end{feynman}
	\end{tikzpicture}	
\end{equation} 
\caption{The simplest diagram with four external fermion lines. The points of interactions are $x_1', x_2'$.}  %
\label{fig.5}   %
\end{figure}
The above expressions may be applied to the more general cases as presented in \cite{Zhang2022} in a straightforward way, as they form the basic building blocks of more general Feynman diagrams.

\section{Non - renormalization of the Hall conductance by interactions}
\label{app3}

In order to prove the non - renormalization of the QHE conductivity by interaction corrections in our case it is sufficient to prove that there are no interaction corrections to the electron density in the moving reference frame, where the electric current disappears.  Without interactions the electron density is calculated as $\rho_0 =  - i \int \frac{d^2x}{S} \int_p  Tr G_{0,W}(x,p) \star \frac{\partial}{\partial p_3} Q_{0,W}(x,p)
$. Here $p_3 = \omega$ is the Matsubara frequency, while $S$ is the system area. By $\int_p$ we denote $\int \frac{d^3 p}{(2\pi)^3}$. We intend to prove the non - renormalization of this expression in the presence of interactions in the following. 

\subsubsection{The first order}

For the contribution of the first order, let us consider
\begin {eqnarray}\label{K1(0)}
\mathcal{K}_1
&=& i \int_p {\rm Tr} \Big[ \Sigma_{1,W} (x,p) \star \partial_{p_3} G_{0,W}(x,p)\Big] \nonumber \\
&=& i \int_{p,k}
{\rm Tr} \Big[  \gamma^{\mu_1}  G_{0,W}(x,p-k) \gamma^{\mu_2} D^{\mu_1 \mu_2}_W(x,k) \star \partial_{p_3} G_{0,W}(x,p)  \Big] ,
\end{eqnarray}
where $D^{\mu_1\mu_2}_W$ is the Wigner transformation of the photon propagator $D^{\mu_1\mu_2}$. In the main text it depends only on $p$. But here we consider the more complicated case of the inhomogeneous photon propagator which depends also on $x$. However, we require
\begin{equation}
	D^{\mu_1 \mu_2}_W(x,k) = D^{\mu_1 \mu_2}_W(x,-k), \quad 	D^{\mu_1 \mu_2}_W(x,k) = 	D^{\mu_2 \mu_1}_W(x,k)\label{Dsym}
\end{equation}
This condition is satisfied obviously for the free photon propagator taken in one of the conventional gauges.   
The self - energy function $\Sigma_{1,W}$ (with external legs attached) is shown in Fig.\ref{fig.2}(a), whereas
$\mathcal{K}_1$ is given in Fig.\ref{fig.6}.
\begin{figure}[h]
	\centering  %
	\begin{equation}
		\begin{tikzpicture}
			\begin{feynman}
				\vertex (a1);
				\vertex [right=2.0cm of a1] (a2);
				
				\vertex [below=2.7em of a1] (b1);
				\vertex [right=1.0cm of b1] (b2);
				
				\vertex [below=1.5em of b1] (c1);
				\vertex [right=1.0cm of c1] (c2);
				
				\diagram* {
					(a1) -- [half left] (a2), 
					(a1) --  [half right, counterclockwise](a2),
					(a1) -- [photon] (a2), (b2) -- [photon] (c2),
				};
			\end{feynman}
		\end{tikzpicture}	
\end{equation}
\caption{Fermionic bubble with current insertion at leading order}  %
\label{fig.6}   %
\end{figure}
Without the partial derivative $\partial_{p_3}$, the above integral is given by the bubble diagram of Fig.\ref{fig.7}, which has the form
\begin {eqnarray}\label{bubble_1}
\mathcal{C}_1 = i \int_{p,k} {\rm Tr}   \Big[\gamma^{\mu_1}G_{0,W}(x,p-k) \gamma^{\mu_2}D^{\mu_1 \mu_2}_W(x,k) \star G_{0,W}(x,p)\Big]  .
\end{eqnarray}

\begin{figure}[h]
	\centering  %
	\begin{equation}
		\begin{tikzpicture}
			\begin{feynman}
				\vertex (a1);
				\vertex [right=2.0cm of a1] (a2);
				\diagram* {
					(a1) -- [fermion, half left, edge label=\(p-k\)] (a2), 
					(a1) --  [half right, counterclockwise, edge label=\(p\)](a2),
					(a1) -- [photon] (a2),
				};
			\end{feynman}
		\end{tikzpicture}	
	\end{equation}
	\caption{Fermionic bubble without current insertion at leading order}  %
	\label{fig.7}   %
\end{figure}
The diagram without current insertion is considered as the "progenitor" of that with current insertion, which can be explained as follows. If we add in $\mathcal{C}_1$ the derivative $\partial_{p_3}$ to the expression standing in the integral over momentum, then the integral vanishes because of the symmetry of the $4$-momentum pair $(p,p-k)$, periodicity in momentum $p$, and because $D^{\mu_1 \mu_2}_W(x,k)$ obeys Eq. (\ref{Dsym}). Therefore,
\begin {eqnarray}\label{bubble_1_partial}
\int_{p,k}
{\rm Tr}  \ \partial_{p_3} \Big[\gamma^{\mu_1}G_{0,W}(x,p-k) \gamma^{\mu_2}D^{\mu_1 \mu_2}_W(x,k)\star G_{0,W}(x,p)\Big]   =0.
\end{eqnarray}
The derivative $\partial_{p_3}$ produces two similar terms, which can be described by the diagram of Fig.\ref{fig.6}. The differentiation, while inserted inside $\mathcal{C}_1$, gives rise to $\mathcal{K}_1$, and therefore $\mathcal{K}_1$ also vanishes. Now, the exact form of the first order loop correction $\rho_1$ is given by
\begin {eqnarray}\label{current_1st}
\rho_1 &=& i \int \frac{d^2 x}{S} \int_{p,q} Tr  \Big(\gamma^{\mu_1} G_{0,W}(x,p-q)  \gamma^{\mu_2}D^{\mu_1 \mu_2}_W(x,q)\star\frac{\partial}{\partial p_3} G_{0,W}(x,p)\Big).
\end{eqnarray}
This form is similar to the above expression of $\mathcal{K}_1$ with an additional $dx$ integral. Thus, we obtain that $\rho_1 =0$.

\subsubsection{The second order} 

Similarly, in the case of the second order we first consider 
\begin{eqnarray}\label{K2(0)}
\mathcal{K}_2 = i \int_p {\rm Tr} \ \Xi_{2,W} (x,p) \star \partial_{p_3} G_{0,W}(x,p),
\end{eqnarray}
where $\Xi_{2,W} = \Sigma_{2,W} + \Sigma_{1,W} \star G_{0,W} \star \Sigma_{1,W}$ is the second order correction to the propagator, while $\Sigma_{2,W}$ is the simply connected second order correction to the self - energy (i.e. the piece that can not be divided into the other two cutting one of the lines).  It can be shown that similar to the above case $\mathcal{K}_2$ also vanishes. We have contributions from two photon propagators $D_W^{\mu_1 \mu_2}(k_1), D_W^{\mu_1 \mu_2}(k_2)$. The expressions for progenitors can be read off from Fig.\ref{fig.3}(b) and Fig. \ref{fig.8}. For example,  from Fig.\ref{fig.8} we have 
\begin{align}
\mathcal{C}^\prime_2 =  \int_{p,k_1,k_2} {\rm Tr} &\Big[ \gamma^{\mu_1} G_{W}(x,p-k_1)  \gamma^{\mu_3}\circ_2\star G_{W}(x,p-k_1-k_2)\gamma^{\mu_4} \star_2\circ G_{W}(x,p-k_1) \gamma^{\mu_2} \star_1\circ G_{W}(x,p) \Big]\nonumber\\
 &\cdot D^{\mu_1 \mu_2}_W(x,k_1) D^{\mu_3 \mu_4}_W(x,k_2) ,\nonumber
\end{align}
On the other hand, $\mathcal{C}^\prime_2$ followed from Fig.\ref{fig.3}(b) will be 
\begin{align}
	\mathcal{C}^{\prime\prime}_2 =  \int_{p,k_1,k_2} {\rm Tr} &\Big[ \gamma^{\mu_1} G_{W}(x,p-k_1)\gamma^{\mu_3} \circ_2 \star G_{W}(x,p-k_1-k_2)\gamma^{\mu_2}{}\star _1\circ G_{W}(x,p-k_2)\gamma^{\mu_4}\star_2\circ G_{W}(x,p)  \Big]\nonumber\\
\cdot  D^{(1)\mu_1 \mu_2}_W(x,k_1) D^{(2)\mu_3 \mu_4}_W(x,k_2) \nonumber
\end{align}
As previously, if we insert the derivative $\partial_{p_k}$ inside the integral over $p$, then the integration vanishes because of the symmetry between the $4$-momenta $(p,p-k_1,p - k_1-k_2)$, and periodicity in $p$. $\mathcal{K}_2$ contains two groups of terms: the ones that  can be obtained as a  result of insertion of $\partial_{p_k}$ into $\mathcal{C}_2$ (and multiplication by a symmetry factor), and the same operation applied to $\mathcal{C}^\prime_2$. In order to show this we may redefine variables as $p' = p-k_1 - k_2, k_1' = - k_1, k_2' = - k_2$ and renumber indices $\mu_i$.   As a result the contribution of $\mathcal{K}_2$ in the correction of the electron density becomes zero.
\begin{figure}[h]
	\centering  %
	\begin{equation}
		\begin{tikzpicture}
			\begin{feynman}
				\vertex (a1);
				\vertex [right=2.0cm of a1] (a2);
				
				\vertex [below=3.0em of a1] (b1);
				\vertex [below=3.0em of a2] (b2);
				
				\diagram* {
					(a1) -- [quarter left] (a2), 
					(b1) --  [quarter right, counterclockwise] (b2),
					(a2) -- [quarter left] (b2), 
					(a1) --  [quarter right, counterclockwise](b1),
					(a1) -- [photon] (a2), (b1) -- [photon] (b2),
				};
			\end{feynman}
		\end{tikzpicture}	
		%\end{equation}
		%\begin{equation}
	\end{equation}
	\caption{Fermionic bubble without current insertion at next to leading order}  %
	\label{fig.8}   %
\end{figure}
One has for the second order contribution $\rho_2$ to the electron density:
\begin{eqnarray}%\label{current_i2}
\rho_2 = i\int \frac{d^2 R }{S}\int_p Tr \Sigma_{2,W} \star\partial_{p_k} G_{0,W} + %\nonumber \\ 
i \int \frac{d^2 R}{S} \int_p Tr \Sigma_{1,W} \star G_{0,W} \star \Sigma_{1,W}\star
\partial_{p_3} G_{0,W}.\nonumber
\end{eqnarray}
The two terms entering this expression are  
\begin{align}%\label{current_i21}
\rho^\prime_2 =  i \frac{1}{2}\int  \frac{d^2 x }{S}\int_{p,k_1,k_2}\,\partial_{p_3} \,Tr \Bigl[&\gamma^{\mu_1}G_{0,W}(x,p-k_1) \gamma^{\mu_3}\circ_2 \star G_{0,W}(x,p-k_1-k_2) \gamma^{\mu_4} \star_2\circ G_{0,W}(x,p-k_1)\gamma^{\mu_2}\star_1\circ \nonumber\\
&G_{0,W}(x,p)\Bigr] \  D^{(1)\mu_1 \mu_2}_W(x,k_1) D^{(2)\mu_3 \mu_4}_W(x,k_2) \nonumber
\end{align}
and
\begin{align}%\label{current_i22}
\rho^{\prime\prime}_2 = i \frac{1}{4}\int \frac{ d^2 x}{S} \int_{p,k_1,k_2} \partial_{p_3} \,Tr \Big[&\gamma^{\mu_1}G_{0,W}(x,p-k_1)\gamma^{\mu_3}\circ_{2}\star G_{0,W}(x,p-k_1-k_2) \gamma^{\mu_2}\star _{1}\circ  G_{0,W}(R,p-k_2)\gamma^{\mu_4} \star_2\circ \nonumber\\ %\nonumber\\
&G_{0,W}(x,p) \Big]D^{(1)\mu_1 \mu_2}_W(x,k_1) D^{(2)\mu_3 \mu_4}_W(x,k_2).   \nonumber
\end{align}
These two integrals are obtained from $\mathcal{K}_2$ adding an additional $dx$ integral. We, therefore, arrive at $\rho_2 = 0$.

%%%%%%%%%%%%%%%%%%%%%%%%%%%%%%%%%%%%%%%%%%%%%%%%%%%%%%%%%%%%%%%%%%%%%%%%%%%%%%%%%%%%%%%%%%%%%%%%%%%%%%%%%%%%%

\subsubsection{Higher orders}

Let us consider the contribution $\rho_j$ of order $j$ to the electron density $\rho$. This contribution is expressed as:
	\begin{equation}
		\rho_j = -i\frac{1}{(2j)! S\beta}\int d^3 w d^3 z_1 ... d^3 z_{2j}  \langle {\rm Tr} \,   (\bar{\psi}_{z_1}\gamma^{\mu_1} \psi_{z_1} \lambda_{\mu_1}) ... (\bar{\psi}_{z_{2j}}\gamma^{\mu_{2j}} \psi_{z_{2j}} \lambda_{\mu_{2j}}) (\bar{\psi}_w \partial_{\hat{p}_3} \hat{Q}_{0}(\hat{p},w) {\psi}_w)\rangle_{connected} 
	\end{equation}
where $\partial_{\hat{p}_k} \hat{Q}_{0}(\hat{p} - A,w) \equiv - \partial_{A_k} \hat{Q}_{0}(\hat{p} - A,w)$ is defined in terms of the Dirac operator in the presence of the external fields. By the subscript "connected" we mean that in the diagram expansion of the above expression we should take into account only the connected diagrams. The vacuum average is taken with the interactions turned off. Using Wick's theorem, $\rho_j$ can be represented as a sum over Feynman diagrams $\rho_{j(\alpha)}$ (the factor $(2j)!$ in the denominator is cancelled out due to reordering of the set ${z_1, ..., z_{2j}}$):
	\begin{equation}
		\rho_j = \sum_{\alpha \in {\cal I}_j } [\alpha] \label{ExpI},
	\end{equation}
where ${\cal I}_j$ is the set of the diagrams of order $j$ with current insertion, and $[\alpha]$ is the numerical value of diagram $\alpha$. Each configuration of fermionic propagators allows $(2j-1)!!$ different contractions, but disconnected diagrams are excluded. Each diagram containing $N(\alpha)$ internal fermionic loops and one external insertion $\partial Q$, has the form:
\begin{eqnarray}
		[\alpha] &=& - i (-1)^{N(\alpha)} \frac{1}{S \beta} \int  \frac{d^3 x d^3 p}{(2\pi)^3} {\rm Tr} \Bigl(\gamma^{\mu_{01}}G_W(x,p-k_{01})\gamma^{\mu_{02}} {}_{...}\overset{(...)}{\circ}_{...} \star ... G_W(x,p) \partial_{p^3} Q(x,p) G_W(x,p)\Bigr)\nonumber\\
		&&  \Pi_{i=1,...,N(\alpha)} \frac{d^3 x_i d^3 k_i}{(2\pi)^3} {\rm Tr} \Bigl(\gamma^{\mu_{i1}}G_W(x_i,p_i-k_{i1})\gamma^{\mu_{i2}} {}_{...}\overset{(...)}{\circ}_{...}  \star ... G(x_i,p_i)\Bigr)\nonumber\\
		&& \Pi_{i,j}D^{(ij)\mu_{...}\mu_{...}}_W (x_i,k_{ij}) \, \Pi_{ijk} D^{(ijk)\mu_{...} \mu_{...}} (x_i, x_j) 
\end{eqnarray}
In this expression by ${}_{...}\overset{(...)}{\circ}_{...} $ we denote one of the star products ${}_{...}{\circ} $, or, ${\circ}_{...} $, or $\overset{(...,...)}{\circ} $ acting on one of the photon propagators and corresponding ejection/absorption of virtual photon lines ending either inside the given fermion loop or inside the adjacent one. Correspondingly, $D^{(ij)\mu_{...}\mu_{...}}_W (x_i,k_{ij}) $ is a photon propagator connecting two points on the same (the $i$ - th) loop, while $ D^{(ijk)\mu_{...} \mu_{...}} (x_i, x_j) $ is a photon propagator corresponding to the $k$ - th line connecting the $i$ -th and the $j$ -th fermion loops. Notice that in the case of a photon propagator that does not depend on coordinates the above expression is simplified considerably, and all circle products of the form of $\circ_{...}$ and ${}_{...}\circ$ are absent. However, the ones of the form $\overset{(...)}{\circ}$ remain. 
  
The corresponding progenitor diagrams arise from the following expression
	\begin{equation}
		\hat{C}_j = -\frac{1}{(2j-1)! }\int d^3 z_1 ... d^3 z_{2j}  \langle {\rm tr} \,   (\bar{\psi}_{z_1}\gamma^{\mu_1} \psi_{z_1} \lambda_{\mu_1}) ... (\bar{\psi}_{z_{2j}}\gamma^{\mu_{2j}} \psi_{z_{2j}} \lambda^{\mu_{2j}}) \rangle_{connected}. \label{Bsource} 	
\end{equation}
If the Feynman diagrams contain bubbles of $j$ photon lines, we have
	\begin{equation}
		\hat{C}_j = \sum_{\lambda \in {\cal B}_j} \gamma(\lambda)[\lambda] \label{ExpB}
	\end{equation}
where ${\cal B}_j$ is the set of bubble diagrams without current insertion, $[\lambda]$ is the numerical value of diagram $\lambda$, and $\gamma(\lambda)$ accounts for the combinatorial factor from Wick's theorem. The sum over all the topologically
different bubbles is $B_j = \sum_{\lambda \in \mathcal{B}_j} [\lambda]$ which is directly related to $\rho_j$, where 
\begin{eqnarray}
	[\lambda] &=&  -i{(-1)^{N(\lambda)}}  \int \Pi_{i=1,...,N(\lambda)} \frac{d^3 x_i d^3 p_i}{(2\pi)^3} {\rm Tr} \Bigl(\gamma^{\mu_{i1}}G_W(x_i,p_i-k_{i1}) \gamma^{\mu_{i2}} {}_{...}\overset{(...)}{\circ}_{...}  \star ... G_W(x_i,p_i)\Bigr)\nonumber\\
	&& \Pi_{i,j}D^{(ij)\mu_{...}\mu_{...}}_W (x_i,k_{ij}) \, \Pi_{ijk} D^{(ijk)\mu_{...} \mu_{...}} (x_i, x_j).  \nonumber
\end{eqnarray}
Bubble diagrams are classified by the number of loops and vertex configurations. The set of bubble diagrams of (\ref{ExpB}) is composed of diagrams with arbitrary numbers of fermion loops that contain $2j$ vertices. If $i_k$ is the number of vertices in the $k$ - th loop, then we may assume $i_1 \leq i_2 \leq ... \leq i_N; \ i_1 +... + i_N = 2j$. These vertices are connected pairwise in all possible ways via photon lines according to the mapping $P : \{a^{(k)}|k = 1, ..., N; a = 1, ..., i_k\} \rightarrow \{b^{(l)}|l = 1, ..., N; b = 1, ..., i_l\}$ where $a^{(k)}$ labels the $a$ - th vertex of the $k$ - th fermion loop. We define the equivalence relation $J$ as follows  $J : P [ ( [a + M]_{ mod i_k } )^{(k)} ] \sim
P[a^{(k)}]$ for $ M \ \in \ \mathbb{Z}$ and $P(a^{(j[k])}) \sim P(a^{(k)})$ for an invertible function $j[k]$ such that $i_{j[k]} = i_k$. This reduces the pairs $\tilde{\cal B}_j = (\{i_1,...,i_N\},P)$ to ${\cal B}_j =  \tilde{\cal B}_j/J$, and 
\begin{equation}
	\hat{B}_j = \sum_{\bar{\lambda} \in \tilde{\cal B}_j} \frac{s(\bar{\lambda})}{n_1(\bar{\lambda})! ... n_{M(\bar{\lambda})}(\bar{\lambda})! i_1(\bar{\lambda}) i_2(\bar{\lambda})  ... i_{N(\bar{\lambda})}(\bar{\lambda})}[\bar{\lambda}],
	\label{ExpB2}
\end{equation} 
where the combinatorial factor counts the number of different pairs $\tilde{\cal B}_j$ that correspond to the same ${\cal B}_j$. There are $M[\bar{\lambda}]$ groups of loops, within each group there are $n_k$ loops of the same length. Here $s(\bar{\lambda})$ is the number of rearrangements of vertices that do not lead to any change of the function $P$. As typical example Feynman diagram we may take the one presented in Fig. 13 of \cite{Zhang2022} with the bosonic propagators replaced by the photon ones. 

To construct the diagrams for $\rho_j$, we first consider closed fermion loops with $2j$ vertices, then mark one with a cross as in \cite{Zhang2022}. The configurations are represented as:
	$$
	\left(\begin{array}{c}
		\{i_1, ..., i_N\}\\ k\\m
	\end{array} \right)
	$$
where $k$ specifies the loop with the cross and $m$ its position (the fermion propagator segment between two photon insertions) within the loop. Consider the equivalence relation $\mathcal{H}$, according to which the configurations with crosses
at different fermionic segments of the same $k$ - th loop are considered as equivalent. Moreover, configurations with crosses drawn on the $k$ - th and $l$ - th loops are equivalent, if $i_k=i_l$. The obtained configurations of fermionic loops with crosses are now supplemented by the ways to connect vertices by dashes. The set of curvy photonic lines (instead of dashed lines from Fig. 13 of \cite{Zhang2022}) can be described by the map ${\cal P}:\{a^{(m)}|m=1,...,N; a=1,...,i_m\} \to \{b^{(l)}|l=1,...,N; b=1,...,i_l \} $, with ${\cal P}^2=1$. Let $\tilde{\cal I}$ represent the set of tetrads:
	$$
	\alpha = \left(\begin{array}{c}
		\{i_1, ..., i_N\}\\ k\\ m\\ {\cal P}
	\end{array} \right)
	$$
One can define the equivalence relation ${\mathcal J}^\prime$ where configurations of photon propagators differing by cyclic rearrangements of vertices along any fermion loop or relabeling of fermion loops of the same length (except the one with the cross) are equivalent. Thus, the set of diagrams contributing to the electron density is:
	$$
	{\cal I} = \tilde{\cal I}/({\cal J}^\prime \otimes {\cal H}),
	$$
where ${\cal H}$ involves moving the cross to loops of equal length and changing ${\cal P}$ correspondingly. We can express the density as:
\begin{equation}
		\rho_j = \sum_{\bar{\alpha} \in \tilde{\cal I}_j} \frac{1}{n_1(\bar{\alpha})!...n_{M(\bar{\alpha})}(\bar{\alpha})!i_1(\bar{\alpha})...i_{N(\bar{\alpha})}(\bar{\alpha})} [\bar{\alpha}] .\label{ExpI2}
\end{equation}
Here, $u(\alpha)$ counts fermion loops of a given diagram $\alpha$ with $i_{k(\alpha)}$ vertices. Next, define the transformation of a bubble $\lambda$:
\begin{eqnarray}
		&& [X[\lambda]] =  -i{(-1)^{N(\lambda)}} \sum_{l = 1, ..., N(\lambda)}  \int \Bigl(\Pi_{j=1}^{N(\lambda)} \frac{d^3 x_j d^3 p_j}{(2\pi)^3}\Bigr) \, \partial_{p_l^3}\, \Pi_{i=1,...,N(\lambda)}{\rm Tr} \Bigl(\gamma^{\mu_{i1}}G_W(x_i,p_i-k_{i1}) \gamma^{\mu_{i2}} {}_{...}\overset{(...)}{\circ}_{...}  \star ... G_W(x_i,p_i)\Bigr)\nonumber\\
		&& \Pi_{i,j}D^{(ij)\mu_{...}\mu_{...}}_W (x_i,k_{ij}) \, \Pi_{i,j,k} D^{(ijk)\mu_{...} \mu_{...}} (x_i, x_j)  \nonumber\\&=& -i\sum_{l=1...N(\lambda)}(-1)^{N(\lambda)} \int \Big(\Pi_{i=1,...,N(\lambda)}  \frac{d^3 x_i d^3 p_i}{(2\pi)^3}\Big) \nonumber \\
		&& Tr \Bigl(  \gamma^{\mu_{l1}}\partial_{p^3_l}G_W(x_l,p_l-k_{l1}) \gamma^{\mu_{l2}} {}_{...}\overset{(...)}{\circ}_{...}  \dots \star ... G_W(x_i,p_i) + ... + \gamma^{\mu_{l1}}G_W(x_l,p_l-k_{l1}) \gamma^{\mu_{l2}} {}_{...}\overset{(...)}{\circ}_{...}  \dots \star ... \partial_{p^k_l} G_W(x_i,p_i) \Bigr) \nonumber\\&&
		\Pi_{i=1,...,l-1,l+1,...,N(\lambda)}{\rm Tr} \Bigl(\gamma^{\mu_{i1}}G_W(x_i,p_i-k_{i1}) \gamma^{\mu_{i2}} {}_{...}\overset{(...)}{\circ}_{...}  \star ... G_W(x_i,p_i)\Bigr)\nonumber\\
		&&  \Pi_{i,j}D^{(ij)\mu_{...}\mu_{...}}_W (x_i,k_{ij}) \, \Pi_{i,j,k} D^{(ijk)\mu_{...} \mu_{...}} (x_i, x_j).
\end{eqnarray}
Then, the transformation for diagrams weighted by $\frac{1}{s(\lambda)}$ is:
	$$
	{\cal X}[B_j] = \sum_{\lambda \in {\cal B}} \frac{1}{s(\lambda)} [X[\lambda]] = \sum_{\bar{\lambda} \in \tilde{\cal B}} \frac{1}{ s(\bar{\lambda})} \frac{s(\bar{\lambda})}{n_1(\bar{\lambda})! ... n_{M(\bar{\lambda})}(\bar{\lambda})! i_1(\bar{\lambda}) i_2(\bar{\lambda})  \dots i_{N(\bar{\lambda})}(\bar{\lambda})} [X[\bar{\lambda}]].
	$$
This symmetrizes contributions from $\partial_p G$. We express it in terms of the diagrams of $\tilde{\cal I}$:
	$$
	{\cal X}[B_j] = \sum_{\bar{\alpha} \in X[\tilde{\cal B}_j]} \frac{S\beta}{n_1(\bar{\alpha})! ... n_{M(\bar{\alpha})}(\bar{\alpha})! i_1(\bar{\alpha}) \dots i_{N(\bar{\alpha})}(\bar{\alpha})} [\bar{\alpha}] = \beta S \rho_j.
	$$
Here, $X[\tilde{\cal B}_j]$ is the set of the diagrams from ${\cal B}_j$ with the crosses added. Since $\tilde{\cal I}_j = X[\tilde{\cal B}_j]$, the set $\tilde{\cal I}_j$ contains the diagrams from $\tilde{\cal B}_j$ with the crosses. Thus, we conclude that
	\begin{equation}
		{\cal X}[B_j] = S\beta \rho_j.
	\end{equation}
As a consequence of the presence of the derivative with respect to momentum in $X[B_j]$ under the integral, following the same logic as above, we arrive at 
	$$
	{\cal X}[B_j] = \rho_j = 0
	$$
This means that the sum of all diagrams contributing to the electric charge is zero due to the integral over the closed Brillouin zone without boundary.

\bibliographystyle{unsrt}
\bibliography{QHE3.bib}

\end{document}